\documentclass[copyright,creativecommons]{eptcs}
\providecommand{\event}{FBTC 2010} % Name of the event you are submitting to
\providecommand{\volume}{??}
\providecommand{\anno}{20??}
\providecommand{\firstpage}{1}
\providecommand{\eid}{??}
\usepackage{breakurl}        

\usepackage{graphicx}
\usepackage{amssymb}
\usepackage{relsize}
\usepackage{xspace}
\usepackage{subfigure}
\usepackage{amssymb,amsbsy,latexsym,amsmath, euscript,amsfonts}
\usepackage{epic,eepic}
\usepackage[all]{xypic}
\usepackage[english]{babel}

\newtheorem{fact}{Fact}[section]

\newtheorem{theorem}[fact]{Theorem}

\newtheorem{example}[fact]{Example}
\newtheorem{definition}[fact]{Definition}

\newcommand{\inter}[1]{\stackrel{\scriptsize #1}\rightarrow}

\newcommand{\xpl}[3]{\ensuremath{{\cal E}_{{#1}, {#2}} ({#3}) }}
\newcommand{\xpla}[3]{\ensuremath{{\cal E}^{'}_{{#1}, {#2}} ({#3}) }}
\newcommand{\xplu}[3]{\ensuremath{{\cal E}^{U}_{{#1}, {#2}} ({#3}) }}
\newcommand{\xplir}[1]{\ensuremath{{\cal E}_{S ,R} ({#1}) }}

\newcommand{\irule}[2]{\frac{ #1}{#2}}

 \title{A Taxonomy of Causality-Based Biological Properties
 \thanks{This work has been partially supported by the British-Crui Partnership Programme 2009.
 }}

\author{C. Bodei$^1$,
A. Bracciali$^1$,
D. Chiarugi$^2$,
and 
R. Gori$^1$
\\
 \institute{$1:$ Dipartimento di Informatica,  Universit\`a di Pisa, Italy}
 \email{\{chiara,braccia,gori\}@di.unipi.it}
 \institute{$2:$ Dipartimento di Scienze Matematiche e Informatiche, 
 Universit\`{a} di Siena, Italy}
 \email{chiarugi3@unisi.it}
 }
 \def\titlerunning{A Taxonomy of Causality-Based Biological Properties}
\def\authorrunning{C. Bodei, A. Bracciali, D. Chiarugi \& R. Gori }
\begin{document}
\maketitle

\begin{abstract}
We formally characterize a set of causality-based properties of metabolic
networks. This set of properties aims at making precise several notions on
the production of metabolites, which are familiar in the biologists'
terminology. From a theoretical point of view, 
biochemical reactions are abstractly represented as causal
implications and the produced metabolites as causal consequences of the
implication representing the corresponding reaction.
The fact that a reactant is produced is represented by means of the chain of 
reactions that have made it exist.
Such representation
abstracts away from quantities, stoichiometric and thermodynamic
parameters and constitutes the basis for the characterization of our properties.
Moreover, we propose an effective method for verifying our properties based
on an abstract model of system dynamics. This consists of a new abstract semantics
for the system seen as  a
concurrent network and expressed using the Chemical Ground Form [6] calculus.
We illustrate an application of this framework to a portion of a real
metabolic pathway.

%Being simplicity one of the main feature of our abstract model, our approach allows us to
%efficiently capture the class of properties of interest.
%Other proposals tackle more elaborate properties and models, but at the cost of 
%model construction and verification procedures that are more expensive.

\end{abstract}

\section{Introduction}

Understanding the relationships amongst the elements of 
biological interaction networks is a relevant problem in Systems Biology.  In the words of
\cite{K02}, ``diagrams of interconnections represent a sort of static
roadmaps, but what we really seek to know are the traffic patterns,
why such patterns emerge, and how we can control them''. Formal descriptions of 
interconnections and methodologies for performing traffic simulations {\em in silico}
can orientate  {\em in vitro} experimentation.

We focus here on metabolic networks, i.e.~the set of the cellular
biochemical pathways involved in energy management and in the
synthesis of structural components. Biochemical pathways are typically
composed of chains of enzymatically catalyzed chemical reactions and
are interconnected in a complex way. This makes difficult to understand the 
overall emerging behaviour of a network, starting from the detailed knowledge 
of the single reactions.

An interesting issue is  the identification of the parts of a network whose integrity is
crucial for certain functionalities. These ``hot points'' represent
candidate drug targets for repressing undesired metabolic functions
involved in pathological states, such as infectious diseases and cancer
\cite{Clyde,Fatumo}. Several properties characterizing different aspects of the 
network functionalities have been introduced in the biological  literature, 
often with slightly different versions for the same property.
What formal methods can offer is a way to make precise and classify properties, 
too often expressed only at an intuitive level.
 
Since, broadly speaking,  {\em causality} plays a key role in finding
chains of reactions that connect the parts of a network,
we base our understanding of properties in terms of causality relations.
%e.g.~for determining causal correlations among molecules that are not
%apparently correlated.  
Following the approach in \cite{BMC08},  in order to give a formal characterization of causality-based properties,  we interpret 
biochemical reactions as ``logical consequences``,  where the source metabolites {\em can} cause, i.e. produce the target ones.
Furthermore, we adopt the notion of {\em explanation} of a certain metabolite. 
Given a set of reactions and initial conditions, an explanation
represents the chain of reactions, causally dependent one from the another, 
that leads to the metabolite.
Our approach  therefore models the biochemical dynamics, capturing causal dependencies, while 
abstracting away from other aspects, like quantities, stoichiometric and thermodynamic
parameters. 

On top of the causality notion, we formalize several properties from a potentially longer list. 
Beyond the relevance of their biological meaning,  
these properties show how the few and simple ingredients we propose are sufficiently expressive to make precise several common notions, often intuitively used in biology. 
Specifically, the set of properties we present concerns the role and the relations of metabolites and reactions within a metabolic network.

We propose an effective method for verifying the formalized causal properties,
based on the construction \emph{once for all} of an abstract representation  of the dynamics of the biological system. The system is specified as a concurrent network in terms of the  Chemical Ground Form calculus~\cite{Ca08}.
%Of course several other choices of the specification language for the description of the biological network would be possible and somehow our approach can be thought to be parametrical w.r.t.~the chosen language. 
We opted for the CGC for its extreme simplicity and well established theories and techniques, while it is, at the same time, sufficiently concrete to capture our properties. For our verification purposes, we have defined a slightly different semantics from the one in~\cite{Ca08,CGL08}.
%Thus, all the properties can be verified in a simple and extremely efficient way, 
%showing that our 
% is simple but concrete enough to capture them.   
%   
It is worth pointing out that our choice  mainly strives for  simplicity. Other specification languages suitable for biological networks,  e.g.~\cite{cfaBioAmb,cfaBioAmbLDL,newcfaBioAmb,CFS05,GLi&c09}, could have been adopted
as well, some perhaps even more expressive, but generally requiring higher costs 
for model construction and verification procedures.

Overall, we are interested in efficiently evaluating the impact of changes on working
hypotheses, such as the variation of the initial conditions and of the sets of reactions, according to a {\em what-if} strategy. The method we propose is meant to be exploited as a sort of preliminary  {\em in silico} screening, aiming at determining the  most promising  experiments to be carried out {\em in vitro}. Finally, we believe that our  framework should be palatable to
biologists, since it is very close to the biochemical intuition of causality and to the spirit of many informal notions currently in use.   

\smallskip
\noindent
{\bf Related Work.}
Due to recent progress of wet-lab techniques, many
metabolic networks are structurally well characterized and can be
reconstructed for many organisms up to the genome-scale level (see
e.g.~\cite{Palsson}). 
%Understanding the relationships between
%structure and function in these networks is one of the main
%goals in the post genomic era of biology. 
%Still, \emph{in silico} models and
%simulations represent powerful tools to gain further insights. 
However, approaches grounded on dynamical modeling, 
e.g.~Metabolic Control Analysis  or Metabolic Flux
Analysis (see~\cite{Fell97}), may encounter difficulties, mainly
because part of the needed kinetic parameters are not known. In
contrast, structure oriented analysis only requires information about
the topology of the investigated networks, which is often known.
Even though this kind of approach may not provide a detailed knowledge of the dynamics underlying the target phenomenon, it allows key properties of metabolic networks to be
addressed, as demonstrated by the plethora of works in the literature. We mention here~\cite{Schuster} and \cite{Schilling}, where the authors
propose to exploit ``elementary modes'' or ``extreme pathways'' to
perform pathway analysis and to assess structural properties, such as
structural robustness and redundancy. In \cite{Stelling}, a method is
presented that relies on the network structure for predicting
robustness in gene regulation networks, while \cite{Alon07} reviews a group of works in which
recurring patterns of interaction in biochemical
networks (a.k.a.~network motifs) are identified and related to specific behaviors or network
robustness. In \cite{Klamt} a novel method is used to target those nodes whose deletion causes the failure of certain network functionalities.

%%% PA %%%
Process algebras have been often used to abstractly model biological systems as concurrent systems, e.g.~\cite{PRSS04,bio_pi,RPS+04,rccs,kappa,brane,beta1}.
%In particular, $\pi$-calculus \cite{pi} and Ambient Calculus \cite{amb} have been transferred from theoretical computer science setting to the biology setting, where suitable biological versions of them, such as the Biochemical stochastic $\pi$-calculus \cite{PRSS04,bio_pi} and BioAmbients \cite{RPS+04} have been introduced. 
%Also a version of CCS, RCCS \cite{rccs}, that addresses biological issues, has been presented.
%Other calculi have been instead specifically defined for biological modelling, 
%such as $\kappa$-calculus \cite{kappa}, Brane calculi \cite{brane} and Beta-binders \cite{beta1}.
Closer to our approach is the work presented in \cite{CDPB04}, where 
the authors apply a causal semantics of the $\pi$-calculus
\cite{pi} in order to describe biochemical processes. 
We use instead CGF, with a simpler semantics, but suitable for 
establishing the causal dependencies of interest.

Our results are close to those obtained by applying Control Flow Analysis (CFA),
a quite efficient static technique, to process calculi used for modeling biological systems, 
e.g.~\cite{CFA_BIO,cfaBioAmb,cfaBioAmbLDL,newcfaBioAmb,B09}.
In all the cases, an over-approximation of the behaviour of a system is offered.
%Simple analyses does not consider the
%possible consumption of reactants.  
%Having an over-approximation of the {\em exact} behaviour of
%a system, both in the case of static analysis and in our framework,
%means that all those events that the prediction does not include will
%{\em never} happen, while when included, the events {\em can} happen,
%i.e.~they are only possible.
In particular, the analyses presented in 
\cite{cfaBioAmbLDL,newcfaBioAmb}
capture causality information relevant for interpreting biological phenomena, and the authors  propose a formalization of properties, like in our approach.
Temporal and causal properties are also addressed in 
\cite{GLi&c09}, where an Abstract Interpretation Analysis for systems specified in the BioAmbients calculus is used to model the quantities of molecules involved in interactions.

%A similar proposal is included in \cite{shank}, 
%where a qualitative model for the analysis of properties is proposed in the context of signalling networks. 
%It would be interesting to integrate the two.

\sloppy
Our approach also shares some similarities with  BIOCHAM \cite{CFS05} and the Pathway Logic \cite{EKLLT02}. 
BIOCHAM is based on the Biochemical Abstract Machine,
 which offers a formal modeling environment for biochemical processes and qualitative descriptions of these processes. BIOCHAM is based on a rule-centered language for specifying biochemical systems and, differently from our approach, it provides tools for querying
temporal properties expressed in the  Computation Tree Logic.
Pathway Logic uses rewriting logic for modeling biological pathways and for enabling the symbolic analysis on them. In a way similar to ours, biochemical reactions are rendered in terms of rules acting on molecules.
%Still close to our approach and also to BIOCHAM is Pathway Logic (see
%e.g.~\cite{EKLLT02,TEKLL04}), where rewriting logic is used for
%modeling biological processes.
%Rewrite rules describe, on the right-hand side, the local change obtained when an instance of the left-hand side is present in the modeled system.
%Rewrite rules describe local changes and the molecular patterns that
%cause them.  
%Logical inference and analysis techniques are exploited for simulation
%to study possible ways a system could evolve.
Both these approaches allow biochemical networks to be specified at a high level of abstraction. However, some expressible features, e.g.~the distinction among different classes of molecules or reactions, have appeared too detailed for the aim of tracking causality and for our quest for a skeletal language for characterizing causality-based relevant properties.
%
%Important biological queries can be asked about the role of a given metabolite, i.e.~whether it is necessary or not for the studied metabolic pathway to be followed.  
%%Refraining from dealing with quantities, BIOCHAM offers explicit
%%controls on reactant consumption during reactions and, by default, all
%%the possibilities are considered.  
%This suggests further developments
%for our approach, where, currently, only the case of no consumption is
%admitted.  The reason for this choice is effectiveness: in this way,
%branching semantics is avoided.
%
%BIOCHAM rules abstract from quantitative parameters in the same way we do: 
%only the presence or the absence of reactants are taken into account.
%Nevertheless, in modelling reaction rules we resort to
%a further abstraction.
%Since the species on the right-hand of a reaction
%may be non-deterministically present or consumed,
%we choose to never exclude the possibility that they can be present.
%What we obtain is therefore more approximate than the set of behaviour considered
%in BIOCHAM. There dynamic evolution of systems is instead 
%taken into account and the lack of knowledge on the possible consumption
%is solved, by considering all combinations of consumption of the reactants, 
%including no consumption at all. 
%
%Rules can be concurrently applied and this corresponds to
%the actual possibility of biological compartments to independently
%evolve.  This offers a basis for
%advanced forms of symbolic analysis.
%At present, in our framework, the concurrent aspect is deliberately ignored.
%

As discussed, several of the mentioned approaches may provide more detailed models and properties than ours, however they generally require computationally expensive verification techniques. Our proposal combines the formalization of properties with a light-weight, approximate in some regards, computational machinery.

%%%%As we already discussed, many of the cited approaches are able to address more complex properties.
%%%%Nevertheless the verification procedures are more expensive and not always efficient.
%%%%Under this regard, our approach
%%%%being based on a more skeletal and abstract
%%%%setting offers efficient ways to verify properties.

\noindent
{\bf Synopsis.} The metabolic network model is illustrated in $\S 2$, properties are formalized in $\S 3$ and the process-algebraic computational framework is introduced  in $\S 4$. An example is discussed in  $\S 5$.
%before some final remarks ($\S 6$). 
%Related work is discussed
%throughout the paper.

\section{A formal model of metabolic networks}
We give an abstract representation of metabolic networks 
and of the corresponding biochemical reactions.
More precisely, we abstract away from quantities,
stoichiometric proportions, kinetic or thermodynamic parameters,
that are involved in reactions, e.g.~consider a standard biochemical reaction like:
\begin{eqnarray}
  aA + bB  
&\rightarrow^{r}& 
  cC + dD
\label{z}  
\end{eqnarray}

\noindent
where $A, B, C$ and $D$ are the species involved, $a, b, c$ and $d$ are the
 corresponding stoichiometric coefficients, and $r$ represents the rate
 at which reactants 
 % eventually
become products.
%
%Since we are interested in investigating causality relationships only,
% we opt to omit the description of many of the factors cited above.  
 We abstractly represent $(\ref{z})$ as:
\begin{eqnarray}
  A \circ B  
&\rightarrow& 
 C \circ D
\label{y}  
\end{eqnarray}
We focus on the fact 
that the presence of both $A$ and $B$ represents
the {\em possibility} for $C$ and $D$ to be $produced$ or $caused$.
Furthermore, we abstract from the dynamic evolution of the network, implicitly assuming that
reactants are never consumed, that it is also
 also an abstraction over their quantities.
 As a consequence $(\ref{y})$ reads as $A \circ B \rightarrow A \circ B \circ  C \circ D$.
 Our model gives therefore an over-approximation of the set
 of the actual pathways, possibly including some pathways that could
 be actually prevented, for instance, by the lack of a suitable
 quantity of reactants or by an inadequate temperature.

For easing the computational machinery, we further decompose any rule
causing more than one metabolite into a set of rules with
only one caused metabolite each, e.g.~rule~$(\ref{y})$ becomes %
$
A \circ B \rightarrow C  \ \ (3)$ plus $A \circ B \rightarrow  D \ \ (4).
$
\noindent
This transformation 
does not impact on causality: the set of metabolites producible by the original
rule can be still produced by applying the new 
rules, as the premises are the same.

%We consider rules with a simplified format, preserving the intended
%meaning and easing the computational machinery. 
Finally, following \cite{Gill00}, the unlikelihood of
reactions involving more than two species, leads us to address only reactions with two
causing metabolites at most. 
%consideration that an interaction involving more than two species at
%the very same time point has a probability close to zero, rules may
%have two causing metabolites at most. 
By summarizing, in the reactions we consider, either two molecules 
%of different species 
produce a new molecule,
or a molecules degrades to another one.

\begin{definition}[{\bf Rules}]\label{bho}
Given a finite set of {\em metabolites} ${\bf M}$, ranged over by
 over by 
$A$, $A_i$, $B$, $C$, $D$..., a
{\em rule} is either in the form (1)
$A_1 \circ A_2 \rightarrow C$, or (2) $A  \rightarrow C$
\end{definition}

The description of causal relations within a metabolic network can
be obtained by defining a set of reaction rules $R$ that describe
how new metabolites can be produced, and  
a set $S$ of metabolites, initially present in the
network solution, which can be seen as premise-less rules.

\begin{definition}[{\bf m\_network} and Initial Solution]\label{sol}
%Given a finite set of {\em metabolites} ${\cal M}$, ranged over by 
%$M,N,P,Q...$, an
An {\em m\_network} $R$ is a finite set of rules with non-empty
premises. 
An {\em initial solution} $S$ is a finite set of premise-less rules in the form 
$\;\;\rightarrow A$.
\end{definition}

The fact that a metabolite is caused by a network is made precise by
means of the following definition relating the metabolite to a chain
of reactions that produce it.

\begin{definition}[{\bf Explanation}]\label{E:1}
$\xpl{S}{R}{C}$ is an {\em explanation} for $C \in {\bf M}$ with respect to $S$ and $R$ if either
\begin{itemize}
\item $C\in S$ and $\xpl{S}{R}{C} = C[~]$, or
\item $A_1 \circ A_2 \rightarrow C = r \ \in\ R$,  
$\exists \xpl{S}{R}{A_1}, \xpl{S}{R}{A_2}$
and $\xplir{C} = C_{r}  [\xplir{A_1}, \xplir{A_2}]$, or
\item $A \rightarrow C = r \ \in\ R$, 
$\exists \xpl{S}{R}{A}$
and $\xplir{C} = C_{r}  [\xplir{A}]$.
\end{itemize}
\end{definition}
Note that  a
metabolite can be initially present in the solution or be
produced anew from the network, or both. These cases can be
distinguished by the structure of the relative explanations.
For simplicity, hereafter in the following definitions we only report the case
of rules in the form $A_1 \circ A_2 \rightarrow C$, by leaving out 
the simpler case of rules is in the form $A \rightarrow C$, where
$\xplir{C} = C_{r}  [\xplir{A}]$.
For observing the explanation structure, we resort to the following auxiliary definition.
% according
%to the following definition.

$
$

\begin{definition}
Given an explanation $\xpl{S}{R}{C}$,
\begin{itemize}
\item
the {\em  set of metabolites} required for $C$,
written ${\cal M}(\xpl{S}{R}{C})$, is defined as follows:

$
{\cal M}(\xpl{S}{R}{C})\ =
\left\{\begin{array}{ll}
C& \mbox{ if } \xpl{S}{R}{C} = C[~]\\
\{A_1,A_2\}\cup \ 
{\cal M}(\xpl{S}{R}{A_1}) \cup  {\cal M}(\xpl{S}{R}{A_2}) & \mbox{ if }
\xplir{C} = C_{r}[\xplir{A_1},\xplir{A_2}]
\end{array}
\right.
$
\item
the {\em  set of reactions} required for $C$, 
written ${\cal R}(\xpl{S}{R}{C})$, is defined as follows:

$
{\cal R}(\xpl{S}{R}{C})\ =
\left\{\begin{array}{ll}
C& \mbox{ if } \xpl{S}{R}{C} = C[~]\\
\{r\}\cup \ 
{\cal R}(\xpl{S}{R}{A_1}) \cup  {\cal R}(\xpl{S}{R}{A_2}) & \mbox{ if }
\xplir{C} = C_{r}[\xplir{A_1},\xplir{A_2}]
\end{array}
\right.
$
%\begin{itemize}
%\item ${\cal R}(\xpl{S}{R}{C})\ =\emptyset$ if   $\xpl{S}{R}{C} = C[~]$, and 
%\item ${\cal R}(\xpl{S}{R}{C})\ = \{r\} \cup {\cal R}(\xpl{S}{R}{A_1i}) \cup  {\cal R}(\xpl{S}{R}{A_2})$ 
%if $\xplir{C} = C_{r}[\xplir{A_1},\xplir{A_2}]$.
%\end{itemize}
\end{itemize}
\end{definition}

%%%\begin{definition}A metabolite $P$ is {\em produced} by a m\_network $R$ in the environment $S$, %%%written $S, R \rhd P$, iff an explanation$\xpl{S}{R}{P}$ exists.\end{definition}

Of course, given $S$ and $R$, there might be more explanation for the same metabolite $C$ 
that corresponds to different ways to produce it.
In turn, the explanation of another metabolite $D$ that uses the metabolite $C$ more than once, 
could include different explanations for $C$ at different points.
For the sake of simplicity, we assume to use only one explanation for each metabolite inside another explanation.
For this reason, we introduce the notion of a {\em uniform explanation}.    
\begin{definition}[{\bf Uniform  Explanation}]
An explanation $\xpl{S}{R}{C}$ for $C \in {\bf M}$ w.r.t. $S$ and $R$ is a {\em uniform  explanation}
(written $\xplu{S}{R}{C}$)
if it is an explanation for  $C \in {\bf M}$
%  according to Definition \ref{E:1} 
and $\forall D\in {\cal M}(\xplu{S}{R}{C})$, if $\xplir{D}$ and $\xpla{S}{R}{D}$ occur in $\xplu{S}{R}{C}$,
then $\xplir{D} = \xpla{S}{R}{D}$, i.e.~$\xplu{S}{R}{C}$ 
does not contain two different explanation for the same metabolite $D$.  
\end{definition}

Since we are going to observe the whole set of explanations for each metabolite, in order to characterize our properties, we do not loose generality, by restricting ourselves to uniform explanations.
From now on, we will then use only uniform explanations and therefore we will omit the superscript $U$.
The following result relates general and uniform explanations. 
\begin{theorem}
Given $S$ and $R$, we have that
$\exists \xpl{S}{R}{C}$ iff $\exists \xplu{S}{R}{C}$ for all $C\in {\bf M}$.
\end{theorem}

%%%%%%%%%%%%%%%%%%%%%%%%%%%%%%%%%%%%%%%%%%%%%%%%%%%%%%%%%%%%%%%%%%%%%%
%%%%%%%%%%%%%%%%%%%%%%%%%%%%%%%%%%%%%%%%%%%%%%%%%%%%%%%%%%%%%%%%%%%%%%
\section{Causality-based properties}
\label{sec:properties} 

Several properties regarding metabolic networks, which are widely
accepted at an informal level, can be made precise within our
framework. Distinguishably, reasoning in terms of
explanations adds an extra level of detail to the definition of the
properties of interest, as well as having an explicit characterization
of the network environment allows us to take  into consideration the
different conditions under which a network may work.  We present 
properties that can be interpreted in terms of our notions of
causality and explanations and that,
given the abstraction of our model, are qualitative properties.
We group them in properties
about reactions and about networks.
The first ones allow us to interpret the results of perturbative
experiments, due to variations of the initial solution $S$ or of the rules in 
$R$, while network properties have to do with robustness.

\paragraph{Reaction properties}

Often, the rules defining the reactions of a metabolic network 
correspond to enzymes that catalyze such reactions or to 
genes that code for such enzymes or for the proteins involved in reactions. 
Rules are hence the main
object when studying a network behavior and it is quite natural
trying to characterize their role in the production of metabolites.
The next definition states when a rule has to be considered {\em
essential} for the production of a given metabolite. 
A rule is essential if it is not dispensable, i.e.~the network, deprived of 
it (e.g.~by knocking-out the corresponding gene), is not able to produce the metabolite. 
%In this context, the network environment is fixed and
%hence the property accounts for a weak form of essentiality. 
%Note that
%taking into consideration the structure of the explanation, viz., its
%length, allows us to abstract away from the fact that the metabolite
%could already be present in the solution and, rather, it is
%required that the network is capable of synthesizing the metabolite.
Generally, in the biological literature, the notion of essentiality has been
expressed informally
and often referred to the elusive 
notion of viability of an organism, e.g.~ \cite{Gerdes}. 

%, which is itself hard to define and
%rather arbitrary
 
\begin{definition}[Essentiality]
Given $R$, a rule $r\in R$ is {\em essential} in $S$ for the metabolite $C$
iff $\exists \ \xpl{S}{R}{C}$ and
$\not\exists \ \xpl{S}{R\setminus r}{C}$.
\end{definition}
Note that if a rule $r$ is essential in $S$ for the metabolite $C$, then
all the explanations of $C$ use $r$.
%for all $\xpl{S}{R}{C}: {\cal R}(\xpl{S}{R}{C}) \ni r$.
{}From a biological point of view, it can be significant to distinguish
amongst two degrees of essentiality. 
In the first case, essential rules correspond to those reactions whose essentiality holds only in a 
given solution $S$. Characterizing these~``hot points'' in a biochemical network operating in a
given solution, can be useful when the studied networks are
typically resident in a well defined environment. This is the case,
e.g.~, of drug development for cancer therapy, as malignant
cells typically live in human blood or inter-cellular matrix. 
Essential rules in the metabolic network of malignant cells represent potential targets for anti-cancer drugs designed for disrupting that network. 
Since cancerous cells always act in a unique environment, it is important to identify their ``weak points'' always considering an initial solution $S$ resembling the composition of human blood or inter-cellular matrix.
In contrast, when the target system is an organism capable of living in
various environments (such as a bacteria), identifying a stronger kind of essentiality, where
a rule is essential for all possible solutions $S$, 
turns out to be a better choice in order to find ``universal'' targets for inhibiting the production of a given metabolite. 
Note that verifying this second kind of essentiality for a certain metabolite $C$ is straightforward, because it simple amounts to verifying 
whether there is only one rule (not having $C$ in the premise) for producing $C$.

Also relationships between rules have been traditionally explored, as has been done with
the notion of mutual essentiality, e.g.~\cite{Yu}.
We say that two rules are {\em mutually essential} for $C$, when 
their individual exclusion does not prevent the production of $C$, i.e.~neither 
of the two rules is essential, but their simultaneous exclusion does. 
Detecting mutually essential reactions can be useful, again, in drug research for identifying multiple 
targets for drugs against parts of a network that represent functional alternatives for the production of a given metabolite.

\begin{definition}[Mutual essentiality]
Given $R$, the rules $r1, r2\in R$ are {\em mutually essential} in $S$
for the metabolite $C$ iff 
$\exists \
\xpl{S}{R\setminus r1}{C}$ and
$\exists \
\xpl{S}{R\setminus r2}{C}$,
while
$\not\exists\ \xpl{S}{R\setminus\{r1,r2\}}{C}$.
\end{definition}

Moreover, we establish that 
two explanations for a metabolite $C$ are {\em vicarious} when
they use different sets of rules, thus representing
two different ways of producing $C$.

\newcommand{\dxpl}[3]{\ensuremath{{\cal E}^{'}_{{#1}, {#2}} ({#3}) }}

\begin{definition}[Vicariate]
Given $R$, and $S$, and a metabolite $C$, an explanation $\xpl{S}{R}{C}$ is {\em vicarious} 
of $\dxpl{S}{R}{C}$ iff ${\cal R}(\xpl{S}{R}{C}) \neq {\cal R}(\dxpl{S}{R}{C})$. 
\end{definition}
\noindent 

This property is related to the previous one, e.g.~if two rules $r_1$ and $r_2$ are
mutually essential for $C$, then 
$\xpl{S}{R\setminus r1}{C}$ is vicarious of $\xpl{S}{R\setminus r2}{C}$.
% when considered explanations in $R$.

Furthermore, we investigate the order in which different metabolites are produced, and in particular 
we determine
 whether the production of a metabolite is a necessary condition 
(i.e.~it is a {\em checkpoint}) for the production of another one. 
\begin{definition}[Checkpoint]
Given $R$ and an initial solution $S$, $B$ is {\em necessary} for $C$ iff 
for all explanations of $C$ $\xpl{S}{R}{C}$,  $B\in {\cal M}( \xpl{S}{R}{C})$. 
\end{definition}
Identifying checkpoints offers some insights on the structure of metabolic networks.
From a topological point of view, checkpoint elements can be related to ``bottlenecks'' 
in molecular interaction networks \cite{YK07}. 
As shown in \cite{YK07} these elements, due to their strategical position in the network, are candidate for being essential as well as the reactions through which they are produced. 

Similarly to the previous property, 
one can be interested in the order between rules and 
whether the application of some rules of $R$ it is a necessary condition for the application of other rules.
\begin{definition}[Causality]
Given $R$ including $r_1$ and $r_2$, and an initial solution $S$, let the metabolite $C$ be the conclusion of rule $r_2\in R$. 
The rule $r_1$ {\em causes} $r_2$ ($r_1\sqsubseteq r_2$) iff for all explanations  $\xpl{S}{R}{C}=C_{r_2} [\xplir{A_1}, \xplir{A_2}] $, either $r_1\in {\cal R}(\xpl{S}{R}{A_1})$ or $r_1\in {\cal R}(\xpl{S}{R}{A_2})$.
\end{definition}
Note that if $r_1\sqsubseteq r_2$ then the metabolite produced by rule $r_1$, say $A$, is necessary for the metabolite produced by rule $r_2$, say $B$, while if $A$ is necessary for $B$ it can be the case that $r_1\not\sqsubseteq r_2$.
Also this property can be exploited (eventually synergically with the checkpoint property) to gain topological insights concerning the investigated network. For instance if a rule $r$ causes a group of other rules it is possible to say that $r$ acts as a bottleneck.

%Roughly, it requires determining the minimal
%environments in which the metabolite can be produced, and checking
%that the rule remains essential throughout the chain leading to the
%more complete environment (monotonicity may help here: if an
%explanation does not exists for the top environment, it will not exist
%throughout the chain). This process can be driven by an expert,
%exploiting domain knowledge.

%DECIDABILITY COMMENT
%Also this property is relative to a fixed solution and a given
%metabolite and is decidable with the same computational cost of 
%essentiality. 

%Now, we consider some properties related to modifications of the initial solution $S$. 
%Note that a certain metabolite could belong to the initial solution $S$, 
%and not being necessary for the production of $C$.

The next property is useful to reason on which metabolites can be omitted from the  initial solution, without compromising the initial capability of the system to produce metabolites in many different ways. Roughly 
speaking a metabolite can be omitted from the initial solution because it not necessary in the production of a given $C$ or it is necessary but the system is  already able to produce it.

Identifying these metabolites can aid in metabolic engineering \cite{TN03}, e.g.~when for optimizing resources usage is requested to characterize the minimal environment needed for a bioreactor. 
Note that the so-called conditional mutants differ from the \emph{wild type} (i.e.~the microorganism possessing the genome commonly found in nature) only for the minimal environment needed for their viability. The genome of conditional mutants do not code for an enzyme essential for its life and their survival is conditioned by the presence in $S$ of the metabolite produced by the missing reaction.  
% without 
%compromising the ability of the system of producing a metabolite $C$ in exactly the same ways as before. 

\begin{definition}[Redundancy]
Given $R$  and a metabolite $C$, an initial solution $S$ is {\em redundant} for $C \not\in S$ iff 
there exists at least a metabolite $B \in S$  
s.t.~for all $\xpl{S}{R}{C}$ for $C$, there exists $\xpl{S\setminus \{B\}}{R}{C}$ such that $  {\cal R}
(\xpl{S}{R}{C})\subseteq   {\cal R} (\xpl{S\setminus \{B\}}{R}{C})$ .
\end{definition}

According to this definition, a given initial solution $S$ can be redundant for a metabolite $C$ in the \emph{wild-type} but not redundant for the same metabolite in the conditional mutant.
Of course, the previous property can be weakened obtaining another property that checks whether $R$ is still able to produce 
$C$ after the exclusion of the reagent $B$ from the initial solution. 
In other words, it addresses the impact of the exclusion of some metabolites from the initial solution, offering straightforward applications to the resource optimization problem described above.

\begin{definition}[Exclusion]
Given $R$, a metabolite  $B\in S$  {\em cannot be excluded} for the production of metabolite $C \not\in S$ iff 
$\not\exists \ \xpl{S \setminus \{B\}}{R}{C}$.
\end{definition}
%$\not\exists \ \xpl{S \setminus \{B\}}{R}{C}$.
Note that
$B$ cannot be excluded for the production of $C$ if and only if
for all explanations $\xpl{S}{R}{C}$, $B[]$ does occur in $\xpl{S}{R}{C}$.
The previous property is in some way related to the checkpoint property expressed before: if $B$ belongs to $S$ and is not necessary for the production of $C$, then we can harmlessly exclude it from the initial solution. 
However, in general, the two properties do not coincide (see Ex.~\ref{es:ma}).

\paragraph{Metabolic network properties}

Informally, robustness can be defined as the capability of a whole network of resisting to damages.
In the biological literature there is not a common agreement on what
``robustness'' exactly means \cite{Kitanoa}. One of the most used definition
says: ``robustness is a property that allows a system to maintain its
functions against internal and external perturbations.''
(\cite{Kitanob}). A similar definition
%, considered by others more appropriate 
\cite{Stellingj} is also widely used:
``robustness, the ability to maintain performance in the face of
perturbations and uncertainty, is a long-recognized key property of
living systems''. 
Both definitions, however, result to be not well
assessed and therefore open to different possible
interpretations. Moreover,
the notion of robustness is related to the maintenance of a
``function'' or of  ``performance''. Both these concepts subsume
quantitative issues and their exact meaning change depending on the
work considered. 
The uncertainty in definitions makes difficult both evaluating robustness effectively and comparing different networks addressing this property.
A more reliable way to assess this notion consists in considering the qualitative features of the network in hand rather than its quantitative throughput \cite{AJB00,SSEB03}. The notion of network robustness can be linked to the overall error tolerance, seen as the capability of carrying information in spite of local failures which, in turn, depend critically on the topology of network wiring \cite{AJB00}. In our framework this corresponds to evaluate the resistance to failures in terms of the maintenance of the capability of producing a given metabolite.    
This paradigm-shift allows us to propose the following formal notions of robustness.

\begin{definition}[Strong Robustness]
Given two m\_networks  $R_1$ and $R_2$, $R_2$ is {\em strongly more robust
than} $R_1$ in $S$ for $C$, written $R_1 \ll_{S,C} R_2$ iff 
$$\exists \ \xpl{S}{R_1}{C} 
\ \ \ \Rightarrow\ \ \  
\exists \ \xpl{S}{R_2}{C} 
\ \ \ \wedge\ \ \  
{\cal R}(\xpl{S}{R_1}{C}) = {\cal R}(\xpl{S}{R_2}{C}).
$$
\end{definition}

This is quite a strong requirement, accounting to say that all the
rules used for producing $P$ in $R_1$, are present in $R_2$ and can be used as well.
Of course, $R_2$ may also allow more explanations, using different rules.
This consideration leads us to formulate a
weaker property, by requiring that $R_2$ is able to produce the same 
metabolite, without constraints on the rules to be used.

\begin{definition}[Weak Robustness]
Given two m\_networks  $R_1$ and $R_2$, $R_2$ is {\em weakly more robust
than} $R_1$ in $S$ for $C$, written $R_1 \prec_{S,C} R_2$ iff 
$$\exists \ \xpl{S}{R_1}{C} 
\ \ \ \Rightarrow\ \ \  
\exists \ \xpl{S}{R_2}{C} 
%\ \ \ \wedge\ \ \  
%\xpl{S}{R_1}{P} \equiv \xpl{S}{R_2}{P}.
$$
%\noindent where $\equiv$ stands for syntactic equality.
\end{definition}

\section{Verification Methodology}
\label{comp}

%To illustrate our methodology, we describe the computational framework
%that represents the counterpart of the formalization of metabolic
%causality introduced in the previous sections.  

Our methodology is based on the construction of an abstract model of the biological system.
The model is obtained by a new abstract semantics of the system,     
interpreted as a concurrency network and expressed using   
the Chemical Ground Form (CGF) \cite{Ca08} calculus.  
We exploit the notion of path for verifying our properties.
%Since causality is a key notion in concurrency, we have chosen the process algebraic setting, and,
%more precisely, the Chemical Ground Form (CGF) calculus \cite{Ca08}.
The CGF is a fragment of the stochastic $\pi$-calculus \cite{Pr05,PC04}.
Since we abstract away from quantities, we resort to a simplified version of CGF,
in which stochastic features, like action rates, are discarded.
In particular, we abstract away from the version of CGF, presented in \cite{CGL08,GL09}, because we represent
solutions as sets of reagents rather than multisets.
\begin{table} 
%  \hrule
%  \vspace{-0.9ex}
  \[\begin{array}{llll}
    R &::= 0 \mid  A = M, R &  \mbox{Reagents Environment} &  \mbox{(empty, or a reagent and Reagents Env)} \\
    M &::= 0 \mid \pi^{\lambda}.S + M &  \mbox{Molecule}  & \mbox{(empty, or an interaction and Molecule)} \\
    S &::= 0 \mid A|S & \mbox{Solution} & \mbox{(empty, or a variable and Solution)} \\
    \pi & ::= a \mid \overline a \mid \tau &  \mbox{Basic Action} & \mbox{(input, output, delay)} \\
    CGF & ::= (R,S) & \mbox{Chemical Ground Form} & \mbox{(reagent environment with initial Solution)}
  \end{array}\]
%\hrule
  \caption{Syntax of simplified CGF}
   \label{tab:syn-cgf}
   \end{table}
The syntax of CGF is defined in
Tab.~\ref{tab:syn-cgf}. 
%Note that since in this preliminary work  we are not interested in modeling stochastic aspects, therefore for now  the basic actions will  not related to rates. 
We  consider a set of ${Names}$ (ranged over by
$a,b,c, \ldots$), a set  of {\em labels} $\cal L$ (ranged over by
$\lambda, \mu \ldots$), and a set ${Mol}$ (ranged over by $A$,$B$,....)
of {\em variables} (the reagents).
A CGF specification is composed by
a (finite) list of reagent definitions $A_i = M_i$, where $A_i$ is a variable that stands for the name of a chemical species
and $M_i$ is a molecule that describes the interaction capabilities of the corresponding species.
The environment $R$ defines the reagents of a solution $S$.
%We write $R.X$ for the molecule associated to $X$ in $R$ and $X \in R$ to indicate that $X$ is a reagent in $R$.
A \emph{molecule} $M$ 
may do nothing, or may change after a delay (e.g.~because of a molecular decay) or may
interact with other reagents. 
A standard notation is adopted: $\tau$ represents a delay; 
$a$ and  $\overline{a}$ model interaction over a shared channel $a$ 
(the input and output, respectively). 
Together with the reagents definition, a CGF includes a
solution $S$, that represents the initial conditions and is described by a parallel composition
of variables, i.e.~a finite list of reagents.
%A solution $P$ evolves according to the definitions of reagents appearing %in the environment $E$.
%Intuitively, a reagent  of $P$ may change after a delay; or
%two  reagents  of $P$ may synchronize on a
%{channel} $a$ at rate $r$. 
This maps onto the initial solution from Def.~\ref{sol}.

In order to distinguish the actions that participate to a move, we label them.
In a CGF $(R,S)$, $R$
is {\em well-labeled}, if basic action labels 
are all distinct. 
We assume to have well-defined reagents environment and, given $R$, to have a definition for each variable $A$ in $R$ or in $S$.
Moreover, given
a label $\lambda \in {\cal L}$, we
use the notation $R.A.\lambda$ to indicate the process $\pi^\lambda.S$,
provided that $A = \ldots +
\pi^\lambda.S + \ldots$ is the definition of $A$ occurring in $R$.  Finally, given a  CGF $(R,S)$, we denote  with $\top$ the set of all molecules occurring in either $S$ or in the rules of $R$.
In the following, we will use ${Sol}$ for the domain of  sets of reagents.
%
%The evolution of a CGF composed by an
% initial solution $S_{0}\in {\cal S}$ and a given environment $R$ is described by a labeled transition system, where we omit $R$, for simplicity.
There is one transition rule for delay actions and one for synchronizations.
Transition are of the form
$$S \xrightarrow{\Theta,\,\hat{S},X } S' \mbox{ with } S,S', \hat{S} \in Sol, \ X\in Mol,
\  \Theta \in \widehat{{\cal L}}= {\cal L}   \cup ({\cal L}  \times {\cal L})$$
%
%$S, S' \in Sol$, and
%labels are s.t.~$\Theta \in \widehat{{\cal L}}= {\cal L}   \cup ({\cal L}  \times {\cal L})$, $\hat{S}, X \in Sol.$
\begin{itemize}
\item
$\Theta$ reports the label(s) of the basic action(s), which participate to the move, 
\item
$\hat{S} \subseteq S_{0}$ reports the subset of the reagents of the initial solution directly involved in the current move
\item
$ X$ reports the {\em unique} reagent produced by the move (i.e.~by the corresponding reaction).
\end{itemize}

\noindent
The Rule ({\bf Delay}) models the
move of a process ${\tau}^{\lambda}.Q$ appearing in the definition of a reagent $A$. The transition records  
the label $\lambda$, the singleton  $A$ if $A$ belongs to the initial solution $S_{0}$ and 
the reagent produced by the reaction.
The Rule ({\bf Sync}) models the
synchronization between two processes 
$a^{\lambda}.Q_1$ and 
$\overline{a}^{\mu}.0$ occurring 
in the definition of $A$ and $B$, resp.
The transition records  the label pair 
$(\lambda,\mu)$, together with 
the part of $S_0$ used in the rule (i.e.~$\{A,B\} \cap S_{0}$) and the possibly new reagent produced by the reaction.
%\vspace{0.1cm}
%  \hrule
  \vspace{0.09cm}
\[\begin{array}{ll}
(\mbox{{\bf Delay}})\;\;\;\;
\irule{R.A.\lambda = {\tau}^{\lambda}.Q }
{S \xrightarrow{ \lambda,\,\{A\}\cap S_{0},Q} S \cup\{Q\}}
%\vspace{0.3cm}\\
&
(\mbox{{\bf Sync}})\;\;\;\;
\irule{R.A.\lambda = {a}^{\lambda}.Q  \qquad
        R.B.\mu = {\overline {a}}^{\mu}.0}
{ 
{S} \xrightarrow{(\lambda,\mu),\, \{A,B\}\cap S_{0},\,Q  }
S \cup \{Q\}} 
\end{array}  \] 
\vspace{0.09cm}
% \hrule 
%\vspace{0.1cm}

\noindent
We denote with $Tr((R, S_{0}))=({\cal S},\rightarrow,S_{0}, R)$ the {\em labeled transition system} (LTS), obtained, 
starting from the initial state $S_0 \in {\cal S} $, w.r.t.~to environment $R$, and with $Gr((R, S_{0}))$ the corresponding graph.
Since environments are well-labeled, 
%e.g.~basic actions have distinct labels, 
different transitions leaving from the same state carry distinct labels.  

For simplicity, we identify each reaction by a label $ \Theta \in \widehat{{\cal L}}$.
In our model, a reaction $r: A \circ B  \rightarrow D$ can be identified by $(\lambda,\mu)$ and rendered by
the following reagent definitions:

\noindent
$
\begin{array}{ll}
\ \ \ \ \ \ \ \ \ \ \ \ \ \ \ \ \ \ \ \ \ \ \ \ \ \ \ \ \ \ \ \ \ \ \ \ \ \ \ \ \ \ \ \ \ \ \ \ & A =  a^{\lambda}.D\\
\ \ \ \ \ \ \ \ \ \ \ \ \ \ \ \ \ \ \ \ \ \ \ \ \ \ \ \ \ \ \ \ \ \ \ \ \ \ \ \ \ \ \ \  \ \ \ \ & B =  \overline{a}^{\mu}.0
\end{array}
$

\noindent
Given an initial solution $S_0 = \{A,B\}$, the system may perform the transition
$\{A,B\} \xrightarrow{ (\lambda,\mu),\{A,B\},D} \{A,B,D\}$, since  $\hat{S} = \{A,B\} \cap \{A,B\} $
and $S' = S_0 \cup \{ D\}$.
If $A$ and $B$ are involved in other reactions, other actions can be added in their specifications, as in the example below, where
we illustrate our approach on a toy reaction network.

\begin{example}\label{ro1}
Consider the initial solution $S_0 = \{ A,B\}$ and an m-network
$R$, consisting of the rules reported below, on the left-hand side.
{\small
\[\begin{array}{crllc}
(\lambda,\mu ) & A \circ B &\rightarrow& D\\
(\delta, \eta)  & A \circ C &\rightarrow& E\\
(\beta,\gamma)& D \circ  B &\rightarrow& A \\
\xi& D &\rightarrow& C\\
(\psi,\nu)  & D \circ C &\rightarrow& E 
\end{array}
\ \ \ \ \ \ \ \ \ \ \ \
\begin{array}{l}
A =  a^{\lambda}.D+ \overline{b}^{\eta}.0 \\
B =  \overline{a}^{\mu}.0+ d^{\beta}.A\\
D = \tau^{\xi}.{C}+\overline{c}^{\nu}.0+\overline{d}^{\gamma}.0\\
C = b^{\delta}.E + c^{\psi}.E 
\end{array}\]
}
The corresponding CGF specification is above on the right, while the corresponding 
graph is in Fig.\ref{fig366}.
%, where 
%$S_{0}=\{A,B\}$, $S_{1}=S_0 \cup \{D\}$, $S_{2}=S_1 \cup \{C\}$, $S_3= S_2 \cup \{E\}$.
For simplicity, in the presence of multiple self-loops, we 
collapse the self-loop arcs in a single one.   
\begin{figure}
\includegraphics[scale=0.60]{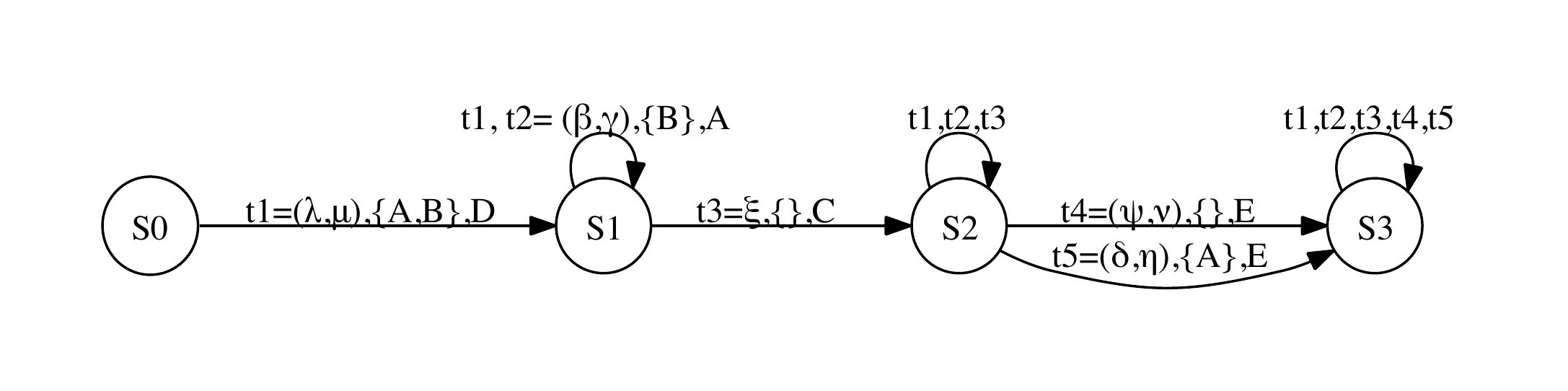}
\caption{LTS Graph of Ex.~\ref{ro1}, where
$S_{0}=\{A,B\}$, $S_{1}=S_0 \cup \{D\}$, $S_{2}=S_1 \cup \{C\}$, $S_3= S_2 \cup \{E\}$
}
\label{fig366}
\end{figure}

\noindent 
$\bullet$
Starting from $S_0 = \{A,B\}$, the only possible transition  (here called $t_1$) is the one that uses  rule $(\lambda,\mu)$ and  leads to the state $S_{1}$ containing $D$. After this, 

\noindent 
$\bullet$ either we can fire transition $t_3$ (rule  $\xi$), that leads to $S_2$, that includes $C$;

\noindent 
$\bullet$
or, we can fire transition $t_2$ (rule ($\beta,\gamma$)) that leads to $S_1$, 
where $A$ is already present. 

\noindent 
$\bullet$
From $S_2$, both transitions $t_4$ (rule  ($\delta, \eta$)) and 
$t_5$ (rule ($\psi,\nu$)) are possible, lead to $S_3$ and produce $E$.

\noindent
$\bullet$
Intuitively, we can observe that some transitions, {\em cause} some others: $t_1$ 
causes $t_2$, $t_3$ $t_4$ and $t_5$,  $t_{3}$ causes both $t_4$ and $t_5$, while $t_4$ and $t_5$ 
are independent from each other.

\noindent
$\bullet$
We have two paths reaching a state that includes $E$:
$S_{0} \inter{t_1} S_{1} \inter{t_3} S_{2} \inter{t_4} S_{3}$
that corresponds to the explanation 
$E_{(\delta, \eta )}[A[],C_{\xi}[D_{(\lambda,\mu)}[A[],B[]]]]$,
and
$S_{0} \inter{t_1} S_{1} \inter{t_3} S_{2} \inter{t_5} S_{3}$,
that corresponds to the explanation $E_{(\psi,\nu) }[D_{(\lambda,\mu)}[A[],B[]],C_{\xi}[D_{(\lambda,\mu)}[A[],B[]]]$.

\noindent
$\bullet$
Establishing which metabolites in $S_0$ are used in each transition, can be useful to investigate their impact on the production of the other metabolites. For instance, $A$ is necessary for the production of $C$ and $E$, because
both the states including $C$ and $E$ are reached, 
using $t_1$ (rule ($\lambda,\mu$)), that requires $A$.
%Finally note that considering transition $t_{2}$ and the information we have recorded on transition $t1$, we can prove  that the presence of reagent $A$ is necessary in the initial solution, since reaction  $(\lambda,\mu)$ (which causes all the other reaction) needs reagent $A$ to be started and reagent $A$ can be produced only \emph{ after} it is needed.  ???

\end{example}
In the following, we are going to make precise the notions only informally introduced in Ex.~\ref{ro1}.
Note that self-loops can
correspond either to the application of a rule already applied or to the application of a rule, that has not already been applied, but that produces a metabolite already present.
Self-loops of the first kind do not add any useful information from a causality point of view, in terms of the properties introduced in the previous sections.
Self-loops of the second kind can be instead useful, especially if we are interested in 
checking the possibility of the system  to produce a certain metabolite even if it  is already present in the initial solution.
%checking whether a metabolite can be excluded by the initial solution, still remaining producible by the network.
We first focus on the computation paths not including self-loops at all, that we call \emph{causally} relevant paths.
This notion is used to verify many properties of the previous section.
%We first formalize the correspondence between paths and explanations. 
%In order to formalize the intuitive ideas of Example \ref{ro1}, we now define a \emph{causal relevant} path on the LTS, i.e., a path that reports the minimal amount of information that allows us to address part of the causality properties described in the previous sections. 
%In order to do this
%it only suffices to consider paths formed by arcs other than self-loops.

\begin{definition}[$\chi$-path]
A path $p$ in $Gr((R, S_{0}))$ is a {\em causally relevant} path ($\chi$-path) if 
$$p = S_{0}\xrightarrow{\Theta_{0},\,  \hat{S}_{0},\,{X_{0}}} S_{1}
\xrightarrow{\Theta_{1},\; \hat{S}_{1},\;X_{1}} S_{2}... S_{m-1} \xrightarrow{\Theta_{m-1},\, \hat{S}_{m-1},\,X_{m-1}}
 S_{m}  \mbox{ and } 
S_{i}\not= S_{i-1} \mbox{ for all } i \in [1,m]
$$
We say that $p$ {\em leads to} $C$ if $C = X_{m-1}$
% = \hat{S}_{m-1}$
(i.e.~if $S_m$ is the first state including $C$).
\end{definition}

\begin{theorem}[Correspondence]\label{120}
Given a $\chi$-path $p$  in $Gr((R, S_{0}))$ that {\em leads to} $C$
%leading to a state including $C$ for the first time, i.e.
\[p= S_{0}\xrightarrow{\Theta_{0},\, \hat{S}_{0},\,X_{0}} S_{1}
\xrightarrow{\Theta_{1},\, \hat{S}_{1},\,X_{1}}S_{2}....S_{m-1} \xrightarrow{\Theta_{m-1},\, \hat{S}_{m-1},X_{m-1}} S_{m}  
 \] 
% such that $C = X_{m-1} 
%such that $C\in X_{m}$ and $\forall i \in I$  $C\not\in X_{i}$, 
%we can obtain the corresponding explanation for $C$ 
% $\xpl{S_0}{R}{C}=tr_p(C)$, using the function  $tr_{\_}(\_)$, which for a given path $p$ and reagent $B$ is defined as follows:
%\begin{itemize}  
%\item $ B[]$, if $B\in S_0$,
% \item  $ B^{r_s}[A^{r_1}_1[X_1],...,A^{r_n}_n[X_n]]$ if $B\in X_j$,
% $\Theta_j$ corresponds to the rule $r_s=A_1\circ  ... \circ A_n\rightarrow B$ and $A^{r_i}_i[X_i]=tr_p(A_i)$.
%  \end{itemize}
let $tr_p()$ be the function, which  
for a given path $p$ and reagent $B$ is defined as follows:
$$
{tr}_{p}(B)\ =\
\left \{
\begin{array}{rcl}
          B_{\Theta}[{tr}_{p}(A_{1}), {tr}_{p}(A_{2})]
        & \hbox{ if } &
        \exists\ i \in [1,m].\ X_{i-1} =  B, \hbox{ and }
        \Theta: A_1\circ A_2\rightarrow B
\\
        &&
        \hbox{ is the rule applied in the transition } t_i,
\\        
        B[ \ ] & \hbox{ if } & B\in S_0.
\end{array}
\right .
$$
We obtain an explanation $\xpl{S_0}{R}{C}$ for $C$, as 
$tr_p(C)$, which uses the same rules of $p$.
 \end{theorem}

Moreover, given an explanation $\xpl{S_0}{R}{P}$ we can construct a set of corresponding paths, starting from 
the subset of the initial solution used in the explanation, i.e.~all the metabolites $A$ such that $A[]$ occurs in $\xpl{S_0}{R}{P}$.
We then proceed by exploring the explanation structure from innermost outermost and therefore firing the transitions corresponding to the rules used in the explanation.
%in  $LTS=Tr((R, \bigcup_{A[]\in E}\{A\} ))$ obtained  by exploring the structure of the proof inside out and applying the same rules in the same relative order. 
Note that, serializing the possible parallelism of the explanation can give rise to a set of paths rather than to a unique path. 
For instance, if we start from an explanation $C_{\Theta_1}[A_{\Theta_2}[B[],D[]],F_{\Theta_3}[E[],G[]]]$, corresponding to the application of rules $\Theta_1 = A \circ F \rightarrow C$, $\Theta_2 = B \circ D \rightarrow A$ and $\Theta_3 = E \circ G \rightarrow F$, then we have two corresponding paths, where the order in which the transitions occur is different.
%Such paths, however, differ for the order in which parallel premises can be produced. 
Note that paths obtained by $\xpl{S_0}{R}{P}$ are $\chi$-paths.
%(each state is different from the others). 
%Similarly, but starting from $S_0$, we can instead construct also paths 
%that may include self-loops and that we call {\em relevant paths}.
%
\begin{definition}[$\rho$-path]
A path $p$ in $Gr((R, S_{0}))$ is a {\em relevant} path ($\rho$-path) if 
$$p = S_{0}\xrightarrow{\Theta_{0},\,  \hat{S}_{0},\,{X_{0}}} S_{1}
\xrightarrow{\Theta_{1},\; \hat{S}_{1},\;X_{1}} S_{2}... S_{m-1}\xrightarrow{\Theta_{m-1},\, \hat{S}_{m-1},\,X_{m-1}}
 S_{m} \mbox{ and } \forall j \in [1,m] \ S_j = S_{j-1} \Rightarrow$$
\[ 
 \begin{array}{lll}
 (i)&X_{j}\in S_0 & \mbox{ (the produced metabolite was already in $S_0$), }
 \\
 (ii)&\{X_{j}\}\cap(\bigcup_{0\leq i<j} X_i)=\emptyset &
 \mbox{ (the produced metabolite was never  produced before), } 
 \\
 (iii)&\{X_{j}\}\cap(\bigcup_{0\leq i<j} \hat{S}_i)=\emptyset & \mbox{  (the produced metabolite was never required before). }
 \end{array}\]
 We say that $p$ {\em leads to} $C$ if $C = X_{m-1}$.
%  = \hat{S}_{m-1}$.
\end{definition}
%Self-loops correspond to the application of rules that produce metabolites already present in the initial solution 
%%The previous definition defines the concept of $R$-relevant path  (here $R$ stands for redundancy) paths. Indeed,  we consider paths formed by  transitions between different states (as before) or by particular  self transition, i.e., transitions representing reactions that introduce reagents originally  present in the initial solution
% (Condition (i)),  not previously produced (see Condition (ii)), and not previously required in the pathway of the reaction (Condition (iii)). 
Intuitively, conditions (i)-(iii) will aid us to determine which 
metabolites could  harmlessly excluded from the initial solution,  identifying the metabolites that  the system itself is able to produce before they are required.  
Note that $\rho$-paths as well as $\chi$-paths are always finite: by definition, a self-loop transition can be included in a $\rho$-path at most once. 
In particular, each $\chi$-path is also a $\rho$-path.

 We are now ready to characterize all the properties introduced in $\S 3$, in terms of $\chi$- and $\rho$-paths.  
 A rule $\Theta$ is {\em essential} for the metabolite $C$ if every $\chi$-path leading to $C$ includes $\Theta$, while
 the  rules $\Theta_1$ and $\Theta_2$ are {\em mutually essential} for $C$ 
 if every $\chi$-path leading to $C$ includes at least one of the two rules.
  \begin{theorem}[Essentiality and Mutual Essentiality]
  
  \begin{itemize}
  \item
  A rule $\Theta$, is {\em essential} in $S_0$ for a reagent  $C \not\in S_0$ iff   
  $\forall$ $\chi$-path  
 $S_{0}\xrightarrow{\Theta_{0},\, \hat{S}_{0},\,X_{0}} S_{1}
\xrightarrow{\Theta_{1},\, \hat{S}_{1},\,X_{1}}S_{2}....\xrightarrow{\Theta_{m-1},\, \hat{S}_{m-1},\,X_{m-1}}
 S_{m}$
 in $Gr((R,S_0))$, leading to $C$,
   there exists at least an $i \in [0,m-1] : \Theta_i=\Theta$.
%  \end{theorem}
%  \begin{theorem}\label{ro:1000}
 \item
  Two rules 
  $\Theta_1$ and $\Theta_2$ are {\em mutually essential} in $S_0$
 for a reagent $C\not \in S_{0}$ iff  
  \begin{itemize}
   \item neither $\Theta_{1}$ nor $\Theta_2$ are essential in $S_0$ for $C$, and
 \item
 $\forall$  $\chi$-path $S_{0}\xrightarrow{\Theta_{0},\, \hat{S}_{0},\,X_{0}} S_{1}
\xrightarrow{\Theta_{1},\, \hat{S}_{1},\,X_{1}}S_{2}....\xrightarrow{\Theta_{m-1},\, \hat{S}_{m-1},\,X_{m-1}}
 S_{m}$  in $Gr((R,S_0))$, leading to $C$,  there exists at least an $i\in [0,m-1]$ s.t.~$\Theta_{i}=\Theta_1$ or $\Theta_{i}=\Theta_2$.
  \end{itemize}
   \end{itemize}
\end{theorem}
%  Moreover, in this context  we can also investigate strong essentiality. 
%   
%  
%   \begin{theorem}
%  The   rule $r$ is  strongly essential for  reagent $C$  if and only if  $r$ is weakly essential  in $\top \setminus C$ for $C$.
%  \end{theorem}

In this context, two $\chi$-paths leading to $C$ represent {\em vicarious} explanations if the 
$\chi$-paths resort to different sets of rules.

  \begin{theorem}[Vicariate]
  Given two $\chi$-paths $p_1$ and $p_2$   in $Gr((R,S_0))$, leading to $C$
  
  $
  \begin{array}{l}
p_1=S_{0}\xrightarrow{\Theta^1_{0},\, \hat{S}^1_{0},\,X^1_{0}} S^1_{1}
\xrightarrow{\Theta^1_{1},\, \hat{S}^1_{1},\,X^1_{1}}S^1_{2}....\xrightarrow{\Theta^1_{h-1},\, \hat{S}^1_{h-1},\,X_{h-1}}
 S^1_{h}
 \\
p_2=S_{0}\xrightarrow{\Theta^2_{0},\, \hat{S}^2_{0},\,X^2_{0}} S^2_{1}
\xrightarrow{\Theta^2_{1},\, \hat{S}^2_{1},\,X^2_{1}}S^2_{2}....\xrightarrow{\Theta^2_{m^2-1},\, \hat{S}^2_{k-1},\,X_{k-1}}
 S^2_{k}\\
 \end{array}
 $
 
 \noindent
    $p_1$ is {\em vicarious} of $p_2$ iff either $h\neq k$ 
    or there exists at least a $j$ 
    s.t.~$\Theta^1_j \not\in \bigcup_{0\leq i<h} \{\Theta^2_i\}$. 
    \end{theorem}

To prove checkpoint properties, 
we exploit the information recorded in $\hat{S}$, in order to check whether a certain metabolite $B$ is {\em necessary}
in the production of a reagent $C$.
 \begin{theorem}[Checkpoint]
 Given $R$ and an initial solution $S_0$,
 reagent $B$ is {\em necessary} for the production of $C$ if
 for all $\chi$-path $S_{0}\xrightarrow{\Theta_{0},\, \hat{S}_{0},\,X_{0}} S_{1}
\xrightarrow{\Theta_{1},\, \hat{S}_{1},\,X_{1}}.... S_{m-1}\xrightarrow{\Theta_{m-1},\, \hat{S}_{m-1},\,X_{m-1}}
 S_{m}$ in $Gr((R,S_0))$, leading to $C$,
then (i) $B\in S_{m-1}$ and  
(ii) if $B \in S_0$, then $B \in (\hat{S}_0\cup ...\cup \hat{S}_{m-1})$.
  \end{theorem}
Conditions (i) and (ii) amount to saying that there is a rule $\Theta_i$ that has $B$ in its premises.

  \begin{example}\label{es:100}
  
Consider the initial solution $S_0 = \{ A,B,D,O\}$ and an m-network
$R$, consisting of the rules reported below on the left, while
the corresponding CGF specification is on the right.
 {\small  \[\begin{array}{crllr}
(\lambda,\mu ) & A \circ B &\rightarrow& C\\
(\delta, \eta)& C \circ F &\rightarrow& P\\
(\beta,\gamma)& D \circ A &\rightarrow& F \\
(\xi, \theta)  & B\circ D &\rightarrow& H\\
(\psi,\nu) & D \circ H &\rightarrow& E \\
(\sigma,\rho)& L\circ O& \rightarrow& H \\
(\phi,\pi) & E\circ H &\rightarrow& L\\
(\omicron,\iota)& C\circ D &\rightarrow& O\\
(\alpha,\zeta) &P\circ O &\rightarrow& E
\end{array}
\ \ \ \ \ \ \ \ \ \ \ \
\begin{array}{l}
A =  a^{\lambda}.C+ \overline{c}^{\gamma}.0 \\
B =  {\overline{a}}^{\mu}.0+ d^{\xi}.H\\
C =  b^{\delta}.P +g^{\omicron}.O\\
D = c^{\beta}.{F}+\overline{d}^{\theta}.0+e^{\psi}.E +\overline{g}^{\iota}.0\\
 E =  h^{\pi}.L\\
F =  \overline{b}^{\eta}.0\\
H = \overline{e}^{\nu}.0+\overline{h}^{\pi}.0\\
L = f^{\sigma}.H\\
O=\overline{f}^{\rho}.0+\overline{l}^{\zeta}.0\\
P=l^{\alpha}.E

\end{array}\]
}
 Figure \ref{fig368} depicts the LTS semantics where $S_{0}=\{A,B,D,O\}$, $S_{20}=\top$, and 
 {\footnotesize
\[ \begin{array}{llllll}
S_{1}=S_0\cup\{C\}&S_{2}=S_0\cup \{F\} &S_3=S_0\cup\{H\}&S_4=S_1\cup\{F\} 
 &  S_5=S_3\cup\{C\} &S_6=S_3\cup \{E\}\\
S_7=S_3\cup \{F\}&S_8=S_4\cup\{P\}& S_9=S_4\cup\{H\} &   S_{10}=S_5\cup\{E\,&S_{11}=S_6\cup\{F\} &S_{12}=S_6\cup\{L\} \\
S_{13}=S_8\cup\{E\}& S_{14}=S_{8}\cup\{H\}&S_{15}=S_{9}\cup\{E\}& S_{16}=S_{10}\cup\{L\} &  S_{17}=S_{11}\cup\{L\} & S_{18}=S_{13}\cup\{H\}\\
S_{19}=S_{16}\cup\{F\}
\end{array} \]}
\begin{figure}
\includegraphics[scale=0.65]{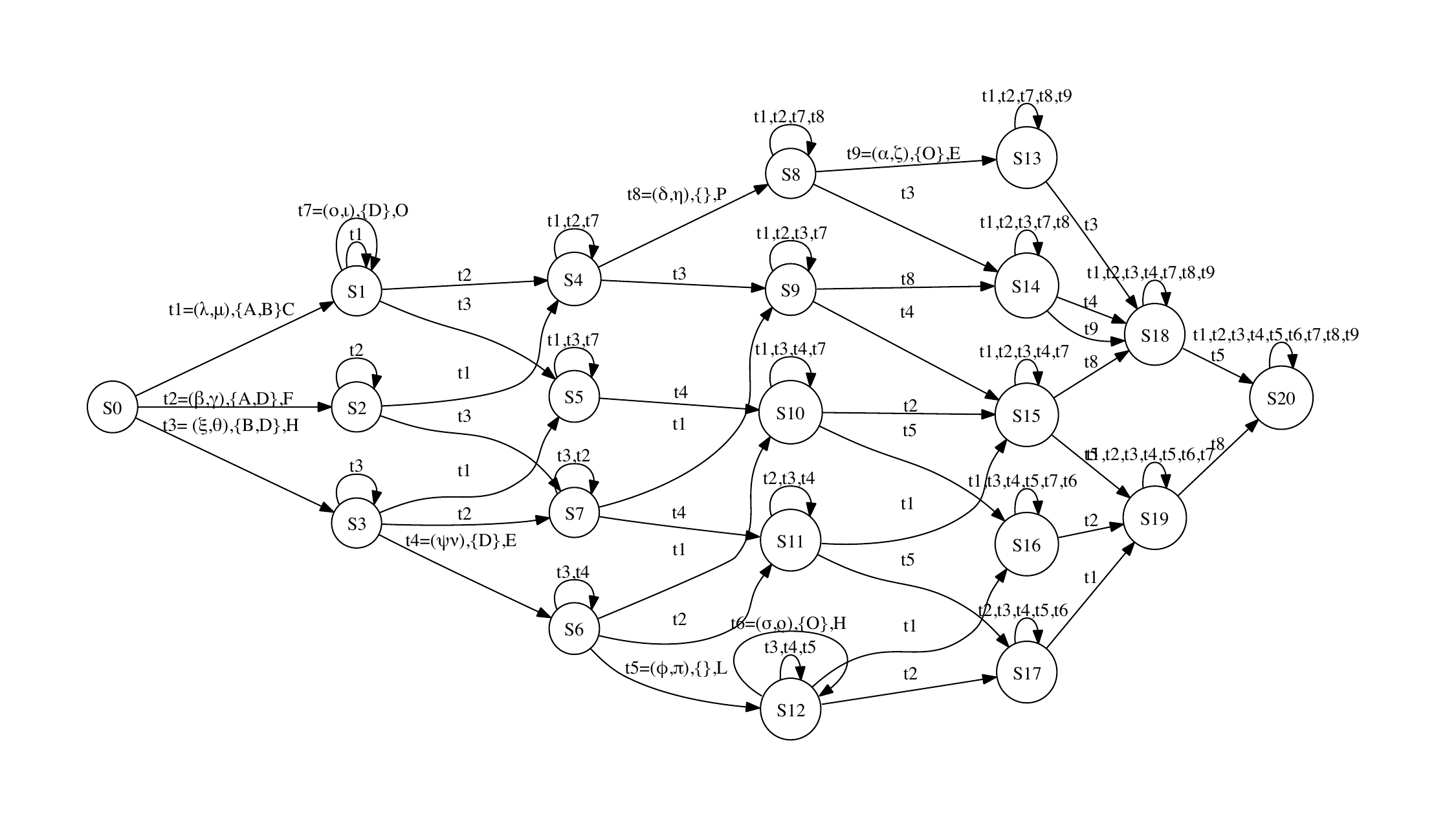}
\caption{LTS Graph of Ex.~\ref{es:100}}\label{fig368}\end{figure} 
We can observe the following properties.

\noindent
$\bullet$  
The production of $H$ is {\em necessary} for that of $L$, 
indeed $H\cap S_0=\emptyset$ and all the states containing $L$
%that are  $S_{12}$ and all  superset of  $S_{12}$  
come after states that include $H$.

 \noindent
$\bullet$  Rule  ($\xi,\theta$) is {\em essential} for the production of $H$.
Actually, also
rule  ($\sigma,\rho$)  is able to produce $H$, but it requires the presence of $L$
%    This can be formally proved by considering all the $\chi$-paths leading to $H$, i.e.~$S_3$, $S_7$, $S_9$, and $S_{18}$.  
%    Intuitively this is because rule (6), which is also able to produce $H$, requires the presence of $L$, 
that in turn requires $H$ to be produced, as discussed before. 
   
%    In our case 
%    the $\chi$-relevant paths leading to   $S_3$ or $S_7$ or $S_9$ or $S_{18}$.  Note that to reach any of these states we have to use  a transition which corresponds to the application of rule (4).  %    However, rule (4) is not strong essential for the metabolite $H$ since 
%    it  is not weak essential for $S_0= S_{20}/\{H\}$. Starting from such an $S_0$ we would obtain $H$ simply applying rule (6).  
 \noindent
$\bullet$  Neither rule $(\psi,\nu)$, nor rule  $(\alpha,\zeta)$ is {\em essential} for the production of $E$, 
because there exists a $\chi$-path $p$ ($p'$) leading to $E$,
 which does not use rule  $(\psi,\nu)$   ($(\alpha,\zeta)$, resp.):
    $p= S_{0}\inter{t_2} S_2\inter{t_1} S_4\inter{t_8} S_8\inter{t_9} S_{13}$, 
    $p'= S_{0} \inter{t_3} S_3 \inter{t_4} S_7$.
 Nevertheless, rules $(\psi,\nu)$  and  $(\alpha,\zeta)$ are {\em mutually essential}. As 
 expected,  
 $p$ and $p'$ represent therefore two alternative ways of producing $E$. 
 The   corresponding explanations:
% Indeed, from $p'$ using Theorem \ref{120}, we obtain the   proof for $E$,
 $E_{(\alpha,\zeta)}[P_{(\delta, \eta)}[C_{(\lambda,\mu )}[A[],B[]],F_{(\beta,\gamma)}[D[],A[]]],O[C_{(\lambda,\mu )}[A[],B[]],D[]]]]$
 (for $p$) and
 $E_{(\psi,\nu) }[D[],H_{(\xi, \theta) }[B[],D[]]]$ (for $p'$)
 represent 
two {\em vicarious} explanations for $E$.

 \noindent
$\bullet$ Note that if we {\em exclude} $A$ from the initial solution, we cannot produce $F$, because
rule $(\beta,\gamma)$ could not be applied,
but we still have a way to produce $E$. 
     \end{example}

  %  rule (5) is not weakly essential for the production of $E$,  since there exists  a
%   $\chi$-path $p$ leading to state $S_13$ containing $E$,
%   which does not use rule (5),
%    $p= S_{0} \stackrel{t1}\rightarrow S_1 \stackrel{t2}\rightarrow S_4 \stackrel{t8}\rightarrow S_8 \stackrel{t9}\rightarrow S_{13}$.

 We now characterize the causality and the robustness properties in our computational framework. 
 We say that a rule {\em causes} another rule, whenever the second one is always preceded by the first one.
\begin{definition}[Causality]
Let $\Theta,\Theta'$ two rules used in $Gr((R,S_0))$. The rule $\Theta$ {\em causes} $\Theta'$
$(\Theta \sqsubseteq \Theta')$ in $Gr((R,S_0))$ iff for all $\chi$-path in $Gr((R,S_0))$ 
$p=  S_{0}\xrightarrow{\Theta_{0},\, \hat{S}_{0},\,X_{0}} S_{1}
\xrightarrow{\Theta_{1},\, \hat{S}_{1},\,X_{1}}S_{2}....\xrightarrow{\Theta_{m-1},\, \hat{S}_{m-1},\,X_{m-1}}
 S_{m}$,
 
 \noindent
$(\Theta'= \Theta_j) \Rightarrow \exists  \Theta_i=\Theta  \mbox{ with }i < j.$
\end{definition}

Robustness has to do with the capacity of a network to produce a certain metabolite. 
 \begin{theorem}[Strong and Weak Robustness]
  Given two environments $R_1$ and $R_2$,  
  \begin{itemize}
  \item
  $R_1  \ll_{S,P}  R_2$  for $C$, iff
   for all $ \chi$-path $p\in Gr((R_1,S_0))$ leading to $C$, $p\in Gr((R_2,S_0))$ and leads to $C$.
  \item
  $R_1 \prec_{S,P} R_2$ for $C$ iff
   for all $\chi$-path $p\in Gr((R_1,S_0))$ leading to $C$,
  then there exists a $\chi$-path $p'\in Gr((R_2,S_0))$ leading to $C$.
  \end{itemize}
  \end{theorem}

% \begin{theorem}[Weak Robustness]
%  Given two environments $R_1$ and $R_2$,  $R_1 \prec_{S,P} R_2$ for $C$ iff
%   for all $\chi$-path $p\in Gr((R_1,S_0))$ leading to $C$,
%  then there exists a $\chi$-path $p'\in Gr((R_2,S_0))$  leading to $C$.
%  \end{theorem}

%
%Note that since we are not concerned about quantities, finer notion of causality relations cannot be captured by our model.

Finally, we characterize the redundancy and the exclusion properties. Both are related with the 
role of initial metabolites and the possibilities of the network to produce metabolites, in case of modifications of the initial set.
%capability of the system to produce a metabolite $C$,   in case of exclusion of a metabolite from the initial solution in the process of  production of a metabolite $C$, 
%%or at least its ability  to be able   to still produce $P$. 
%To this aim we resort to  $\rho$-paths and to the following notion, that given a path $p$, computes
%the subset ${\cal U} (p)$ of the initial solution strictly required to perform all the transitions in a given path. 
%It is composed by the union of all subsets $\hat{S}_i $ of the initial solution $S_0$ used in each transition, a part from the metabolites present in $S_0$ that are however produced in the path.
%If indeed some metabolites present in $S_0$ can be produced in one of the transitions, before their use, actually they do not play an essential rule in $S_0$.
To this aim we resort to $\rho$-paths and to the following notion, that given a $\rho$-path $p$, computes
the subset of metabolites ${\cal U}(p)$ of the initial solution strictly required
to perform each transition of the given path. 
Such information is obtained by collecting
all the subsets $\hat{S}_i $ (i.e.~the subsets of the initial solution $S_0$ used by the transitions in $p$), and by removing those metabolites (in $S_0$) that the system itself is able to produce along the path.
%If indeed such metabolites can also be produced in one of the path transitions, we can deduce that their presence in $S_0$ is not strictly necessary.
Recall that since $p$ is a $\rho$-path, we are guaranteed that the transitions that produce these metabolites always come before
the transitions that use them.
%Therefore, if a metabolite in $S_0$ can also be produced in one of the transitions, before they are  used, their presence in  $S_0$ is not needful.

\begin{definition}
Given a  $\rho$-path in $Gr((R,S_0))$, 
$p=S_{0}\xrightarrow {\Theta_{0},\,  \hat{S}_{0},\,{X_{0}}} S_{1}
\xrightarrow{\Theta_{1},\; \hat{S}_{1},\;X_{1}} S_{2}... S_{m-1} \xrightarrow {\Theta_{m-1},\,  \hat{S}_{m-1},\,{X_{m-1}}} S_{m}   $

\noindent
${\cal U} (p)=(\bigcup_{0\leq i<m}\hat{S}_i  \setminus \bigcup_{0\leq i<m}\{X_i\})$
 \end{definition}

An initial solution is {\em redundant} for the production of a metabolite $C$, whenever there exists at least a component that is not required
from the very beginning, in {\em all the paths} that lead to $C$.
Moreover, to produce a metabolite $C$, we can {\em exclude} a metabolite $B$ from the initial solution $S_0$,
if $B$  is not required from the very beginning, 
in {\em at least one path} that leads to $C$.
\begin{theorem}[Redundancy and Exclusion]\label{ro30}
Given an environment $R$, an initial solution $S_0$, and a metabolite $C \not\in S_0$,
let ${\cal P}_C=\{p\;|\; p \ is \ \mbox{${\rho}$-path} \ in \ Gr((R,S_0))\ that \ leads \ to \ C\}$. Then
\begin{itemize}
\item
$S_0$
is {\em redundant} for $C$ iff $ \bigcup_{p\in {\cal P}_C}{\cal U}(p) \subset S_0$.
 \item
 a metabolite $B$
{\em can be excluded} for the production of $C$ iff 
$\exists$   
a \mbox{$\rho$-path $p$ in ${\cal P}_C$} s.t.~$ B \not\in{\cal U}(p) $.
 \end{itemize}
 
 \end{theorem}

%need not to be present from the very beginning.
%\begin{theorem}[Exclusion]
%Given an environment $R$, an initial solution $S_0$, and a metabolite $C \not\in S_0$,
%a metabolite $B$
%{\em can be excluded} for the production of $C$ iff 
%$\exists$   
%a \mbox{$\rho$-path $p$ in $Gr((R,S_0))$ leading to } $C$ s.t.~$ B \not\in{\cal U}(p) $.
%\end{theorem}

\begin{example}\label{es:ma}
Consider again the network described in Ex.~\ref{es:100}. 

 \noindent
$\bullet$        
The initial solution $S_0 = \{A,B,D,O\}$ is {\em redundant} for the production of $E$. 
Consider indeed the $\rho$-path leading to $E$:  $p_1= S_{0} \inter{t_1} S_1 \inter{t_7} S_1 \inter{t_1} S_4\inter{t_8} S_8 \inter{t_9} S_{13}$. 
Note that $ p_1$ is similar to the $\chi$-path $p$, seen in Ex.~\ref{es:100}, except that it also includes the self-loop transition $t_7$ 
(rule  ($\omicron,\iota$)) on the state $S_1$. 
This transition corresponds to a reaction that produces $O$, which is already in $S_0$, but that it is not required until this point. 
Therefore $O$ could safely be {\em excluded} from $S_0$, since $O \not\in {\cal U}(p_1)$. 
Similarly, we can prove that $O\not\in {\cal U}(p)$ for all the other paths that reach $S_{13}$ and all its successors.  

 \noindent
$\bullet$ 
Consider again
the $\chi$-path $p'= S_{0} \inter{t_3} S_3 \inter{t_4} S_7$, leading to  $E$.
Note that  $p'$ is also a $\rho$-path and that $O\not\in {\cal U}(p')$. 
The same result holds for all the paths reaching $S_{13}$ and therefore the successor states. 
Hence, by Theorem~\ref{ro30} we can conclude that the metabolite $O$ could safely be {\em excluded} from the initial solution without compromising the production of the metabolite $E$. 
If we are not interested in maintaining all the ways to produce $E$, but just the general ability of the system to produce it, 
we can {\em exclude} $A$, since $p'$ is a $\rho$-path leading to $E$ and $A\not\in {\cal U}(p')$.
  
\noindent
$\bullet$  
Note that in this case, checkpoint and exclusion properties rely on the same information:
%the information needed for the notions of checkpoint and exclusion coincide: 
we could have detected indeed that $A$ could be {\em excluded} from the fact that $A$ was not {\em necessary} for the production of $E$. However, this is not true in general. Assume, e.g., to modify rule $(\psi,\nu) $ in order to require the presence of $O$, as $ (\psi,\nu)':  \, O + H \rightarrow E $.  As a consequence, also the paths leading to state $S_7$ require the presence of $O$, making also $O$ {\em necessary} for the production of $E$. However, we can conclude that while $S_0$ is not {\em redundant} for the production of $E$ in the modified system, considering $p'$ above, $O$ could be {\em excluded}, since throughout $p'$ the modified system is still able to produce $E$.
  
     \noindent
$\bullet$  
  Finally note that $(\lambda,\mu )\sqsubseteq (\phi,\pi)$, $(\beta,\gamma)\sqsubseteq (\phi,\pi)$ while neither $(\lambda,\mu )\sqsubseteq (\beta,\gamma)$ nor
  $(\beta,\gamma)\sqsubseteq (\lambda,\mu )$. Indeed, the transitions related to the application of rules 
 $ (\lambda,\mu )$ and $ (\beta,\gamma)$ ($t_1$ and $t_2$ resp.) are not causally related, hence, they can be fired in any order.
  \end{example}

\section{Properties at work in a metabolic pathway}
A precise characterization of the structural role played by the single elements in the overall metabolic
networks is relevant both for better understanding living systems and for 
developing treatments for
pathological aspects. As an example consider the clinical studies of
primary and metastatic cancers that have clearly demonstrated that
human malignancies are characterized by an increased activity of
glycolysis when compared to normal tissue \cite{Gambhir}. This
metabolic peculiarity suggests an inviting target for cancer treatment
and various therapeutic strategies aiming at selectively disrupting glycolytic 
network of malignant cells are under investigation~\cite{Gatenby}.

In this light,  we present   
a simplified glycolytic pathway embedded in a wider context comprising also the Penthose Phosphate Pathway. Through these interconnected pathways the $\beta$-D-Glucose-6P is oxidized yielding Pyruvate and energy (ATP) or Ribose and reducing equivalents (NADPH). 
 
The pathway can be formalized as in Tab.~\ref{glyc_path}.
For lack of space we do not show here the corresponding LTS graph, however it should be clear how our properties, related with very important biological features, can be verified
using the method illustrated in $\S$~\ref{comp}  (see in particular  Ex.~\ref{es:100} and \ref{es:ma}).

%\begin{table}
%\begin{center}
%{\small
%\begin{tabular}{|c|} \hline
%\mbox{$
%\begin{array}{lrcl}
%(1) & {\mbox{\em  $\beta$-D-Glucose}} \circ ATP
%&\ \ \rightarrow\ \ \ &
%{\mbox{\em  $\beta$-D-Glucose-6P}} \circ ADP
%\\
%(2) & {\mbox{\em  $\beta$-D-Glucose-6P}}
%&\rightarrow&
%{\mbox{\em  $\beta$-D-Fructose-6P}}
%\\
%(3) & {\mbox{\em  $\beta$-D-Fructose-6P}} \circ ATP
%&\rightarrow& 
%{\mbox{\em  $\beta$-D-Fructose-1,6bP}}  \circ ADP
%\\
%(4) & {\mbox{\em  $\beta$-D-Fructose-1,6bP}} 
%&\rightarrow&
%{\mbox{\em  Glyceraldehyde-3-P}}  \circ Dihydroxyacetone phosphate
%\\
%(5) & {\mbox{\em  Glyceraldehyde-3-P}}  
%&\rightarrow&
%Dihydroxyacetone phosphate
%\\
%(6) & Dihydroxyacetone phosphate
%&\rightarrow&
%{\mbox{\em  Glyceraldehyde-3-P}} 
%\\
%(7) & {\mbox{\em  Glyceraldehyde-3-P}}  \circ NAD
%&\rightarrow&
%1,3 \ Bisphosphoglycerate \circ NADH
%\\
%(8) & 1,3 \ Bisphosphoglycerate \circ ADP
%&\rightarrow&
%{\mbox{\em  3-Phosphoglycerate}}  \circ ATP
%\\
%(9) & {\mbox{\em  3-Phosphoglycerate}} 
%&\rightarrow&
%{\mbox{\em  2-Phosphoglycerate}} 
%\\
%(10) & {\mbox{\em  2-Phosphoglycerate}} 
%&\rightarrow&
%Phosphoenolpyruvate
%\\
%(11) & Phosphoenolpyruvate \circ ADP
%&\rightarrow&
%Pyruvate \circ ATP
%\end{array}
%$} \\ \hline
%\end{tabular}
%}
%\end{center}
%\caption{Rules of the Glycolytic Pathway}
%\label{glyc_path}
%\end{table}

\begin{table}
\begin{center}
{\small
\begin{tabular}{|c|} \hline
\mbox{$
\begin{array}{lrcl}
(1) & {\mbox{\em  $\beta$-D-Glucose}} \circ ATP
&\ \ \rightarrow\ \ \ &
{\mbox{\em  $\beta$-D-Glucose-6P}} \circ ADP
\\
(2) & {\mbox{\em  $\beta$-D-Glucose-6P}}
&\rightarrow&
{\mbox{\em  $\beta$-D-Fructose-6P}}
\\
(3) & {\mbox{\em  $\beta$-D-Fructose-6P}} \circ ATP
&\rightarrow&
{\mbox{\em  $\beta$-D-Fructose-1,6bP}}  \circ ADP
\\
(4) & {\mbox{\em  $\beta$-D-Fructose-1,6bP}}
&\rightarrow&
{\mbox{\em  Glyceraldehyde-3-P}}  \circ Dihydroxyacetone phosphate
\\
(5) & {\mbox{\em  Glyceraldehyde-3-P}} &\rightarrow&
Dihydroxyacetone phosphate
\\
(6) & Dihydroxyacetone phosphate
&\rightarrow&
{\mbox{\em  Glyceraldehyde-3-P}}
\\
(7) & {\mbox{\em  Glyceraldehyde-3-P}}  \circ NAD
&\rightarrow&
1,3 \ Bisphosphoglycerate \circ NADH
\\
(8) & 1,3 \ Bisphosphoglycerate \circ ADP
&\rightarrow&
{\mbox{\em  3-Phosphoglycerate}}  \circ ATP
\\
(9) & {\mbox{\em  3-Phosphoglycerate}}
&\rightarrow&
{\mbox{\em  2-Phosphoglycerate}}
\\
(10) & {\mbox{\em  2-Phosphoglycerate}}
&\rightarrow&
Phosphoenolpyruvate
\\
(11) & Phosphoenolpyruvate \circ ADP
&\rightarrow&
Pyruvate \circ ATP
\\
(12) & {\mbox{\em  $\beta$-D-Glucose}}  \circ NADP^+
&\rightarrow&
\mbox{{\em D-Glucono-1,5-Lactone-6P}} \circ NADPH\\
(13)&
\mbox{{\em D-Glucono-1,5-Lactone-6P}}  &\rightarrow& \mbox{{\em 6-Phospo-D-Gluconate}} \\
(14)&
\mbox{{\em 6-Phospo-D-Gluconate}}  \circ NADP^+ &\rightarrow&  
\mbox{{\em Ribulose-5-P}}  \circ NADPH\\
(15)&
\mbox{{\em Ribulose-5-P}} &\rightarrow& \mbox{{\em D-Xylulose-5-P}}\\
(16)&
\mbox{{\em Ribulose-5-P}} &\rightarrow& \mbox{{\em D-Ribose-5P}}\\
(17)&
\mbox{{\em D-Ribose-5P}} \circ \mbox{{\em D-Xylulose-5P}}&\rightarrow& 
\mbox{{\em Glyceraldehyde-3-P}} \circ \mbox{{\em D-sedoeptulose-7-P}} \\
(18)&
\mbox{{\em D-sedoeptulose-7-P}} \circ  \mbox{{\em Glyceraldehyde-3-P}} &\rightarrow&
\mbox{{\em D-Erythrose-4P}} \circ 
\mbox{{\em D-Fructose-6-P}}\\
 (19)&
\mbox{{\em D-Erythrose-4P}} \circ \mbox{{\em D-Xylulose-5-P}} 
&\rightarrow&
\mbox{{\em Glyceraldehyde-3-P}}  \circ \mbox{{\em $\beta$-D-Fructose-6P}} 
\end{array}
$} \\ \hline
\end{tabular}
}
\end{center}
\caption{Rules of the Glycolytic Pathway and of the Pentose Phosphate Pathway}
\label{glyc_path}
\end{table}

%\paragraph*{Reaction properties: essentiality}
\noindent
{\bf Reaction properties}  
Suppose that our initial solution is $S_{\alpha}$:
$\{
{\mbox{\em  $\beta$-D-Glucose}}$, 
$ATP$, $NADP^+$, 
$NAD\};$
we can verify the following properties.

\noindent
$\bullet$
The existence of the following causal chains of rules:
$(1) \sqsubseteq (2) \sqsubseteq 
... \sqsubseteq (11)$ and
$(12) \sqsubseteq
... \sqsubseteq (15)$.

\noindent
$\bullet$
Rule (7)
is {\em essential} for the production of the metabolite $Pyruvate$.
Its exclusion interrupts all the possible paths reaching $Pyruvate$.

\noindent
$\bullet$
Rule (12) is {\em essential} for the production of various metabolites, 
e.g.~$NADPH$ and $\mbox{{\em D-Erythrose-4P}}$. 

\noindent
$\bullet$
The metabolite $\mbox{{\em D-Glucono-1,5-Lactone-6P}} $ is a {\em checkpoint} for the production
of NADPH, produced both by the rules (12) and (14), and for that of 
$\mbox{{\em D-Xylulose-5-P}} $, produced both by the rules (15) and (17). 
The rule (12) which produces $\mbox{{\em D-Glucono-1,5-Lactone-6P}} $ corresponds to the reaction catalyzed by the enzyme Glucose-6-Phosphate Dehydrogenase(G6PD), an enzymopathy commonly known as fauvism. 

\noindent
$\bullet$
The metabolite $\mbox{{\em Glyceraldehyde-3-P}}$ can be produced either by the $\chi$-path
composed by the transitions corresponding to the rules: (12), (13), (14), (15), (16), (17), or
by the $\chi$-path
composed by the transitions corresponding to the rules: (1), (2), (3), (4).
The two paths correspond to two {\em vicarious} explanations.

\noindent
$\bullet$
The metabolite $NADP^+$ should be {\em included} in the initial solution in order to produce 
$\mbox{{\em D-Xylulose-5-P}}$, but it can be {\em excluded} as far as the production of
${\mbox{\em  $\beta$-D-Fructose-1,6bP}}$ is concerned.

% Recall that a rule $r$ is
%essential for a metabolite $C$, given a solution $S$, iff $C$ can be produced by the network, but there is not an
%explanation for $C$ that does not contain $r$. Given the \mbox{solution $S_{\alpha}$}:
%$\rightarrow
%{\mbox{\em  $\beta$-D-Glucose}}$, 
%$\rightarrow
%ATP$, and
%$\rightarrow NAD$, 
%$ 
%\begin{array}{clcclccl}
%\rightarrow&
%{\mbox{\em  $\beta$-D-Glucose}}
%&\qquad &
%\rightarrow&
%ATP
%&\qquad &
%\rightarrow&
%NAD
%\end{array}
%$

%Rule $(2)$ is not strongly essential since if the metabolite
%$\beta$-D-Fructose-6P is directly added to the initial solution
%$S_\alpha$ obtaining $S_{\alpha}'$, then the metabolite Pyruvate
%will be produced in spite of the removal of the rule,
%i.e.~$\exists \xpl{R\setminus (2)}{S_{\alpha}'}{Pyruvate}$.

%Differently, rule $(11)$ 
%%
%%%%$Phosphoenolpyruvate \circ ADP \rightarrow Pyruvate \circ ATP$ 
%%
%results to be strongly essential for Pyruvate. Indeed, $S_\alpha$ is
%the minimal solution leading to the production of Pyruvate, for
%which $(11)$ is weakly essential. For all the $S$, such that
%$S_\alpha\subseteq I$, the metabolite cannot be produced if $(11)$ is
%removed.

\noindent
$\bullet$
Finally, having the initial solution $S_{\beta} = 
S_{\alpha} \ \cup \ \{ {\mbox{\em  Glyceraldehyde-3-P}}$\},
%$$\
%\begin{array}{rclcrclcrclcrcl}
%&\rightarrow&
%{\mbox{\em  $\beta$-D-Glucose}} 
%&\qquad &
%&\rightarrow&
%ATP
%&\qquad &
%&\rightarrow&
%NAD
%&\qquad &
%&\rightarrow&
% {\mbox{\em  Glyceraldehyde-3-P}}  
%\end{array}
%$$ 
%
the rules (4) 
and
(5) 
are {\em mutually essential} for  $Pyruvate$ as they
must be both removed in order to suppress the
Dihydroxyacetonephosphate production. Another example of mutually
essential rules is given in~\cite{BMC08}.

\noindent
{\bf Network properties}  
In order to illustrate our definition of robustness, we consider two
pathways: the pathway described above and another one,
obtained from the first one by 
suppressing the two 
reactions, one inverse of the other, represented by rules $(5)$ and
$(6)$. This suppression corresponds to the
inhibition of an enzyme, that
is related to a severe disease, known as triosephosphate isomerase (TPI) deficiency, 
see~\cite{Orosz} for details. Considering the standard
solution $S_\alpha$, it is easy to verify that the glycolytic pathway
results to be more robust than its variant related to the disease with
respect to the production of $Pyruvate$: only one of the explanations
for $Pyruvate$ existing in the original network is viable in the second
one. This simple example well highlights the relevance that a study of
robustness may have.
Quite naturally, this notion can be extended in order to consider
robustness with respect to different solutions or with respect to
different metabolites of the same network. About the latter,
intuitively, it turns out that there are at least two explanations for
metabolites like $\mbox{{\em Glyceraldehyde-3-P}}$, 
$\mbox{{\em 2-Phosphoglycerate}}$ and $Pyruvate$,
given the solution $S_\alpha$ in the glycolytic pathway.  Instead,
only one explanation exists for metabolites like $\mbox{{\em $\beta$-D-Glucose-6P}}$
or. Therefore, the network results more
robust for the production of $\mbox{{\em Glyceraldehyde-3-P, 2-Phosphoglycerate}}$
and $Pyruvate$ rather than for that of $\mbox{{\em$\beta$-D-Glucose-6P}}$. From a drug research point of view,
targeting the parts of the network involved in the production of the
last two metabolites may result more effective than targeting the
others. Indeed, a drug targeting the reaction of
hexokinase, leading to the production of $\mbox{{\em$\beta$-D-Glucose-6P}}$ is under
development~\cite{Gatenby}.

\section{Conclusions}

We have presented a taxonomy of biological properties of interest regarding metabolic network. 
Based on a (formal) notion of {\em causality}, this taxonomy translates a bunch of properties in use within biologists into a formal framework. We have also proposed a computational counterpart of the framework, which, allowing the automated verification of the properties, paves the way to the development of software tools supporting the analysis of metabolic networks. 
We have chosen a reading of causality and the computational mechanisms that rely on theories developed in concurrency and particularly suitable to describe causality in interactive behaviours and providing a wealth of analysis techniques. 
Definitions do not depend on the computational framework, and this can be changed whenever another computational support may result more convenient for a specific domain or set of (causally-based) properties.

Future work regards the extension of the set of proposed properties, experimentation with case-studies of interest for wet-lab research, possibly contrasting the framework with other analogous proposals especially as far as the trade-off between expressiveness and efficiency is concerned. Moreover, we would like to attempt a characterization of properties of sets of interconnected signaling pathways, like the ones involved in cancerogenesis, since the understanding of the structural features underlying their interactions may provide useful hints for drug research. In this sense, it could be worth studing possible integrations of our framework with the qualitative logical view adopted in \cite{shank} for signaling networks.

%\section{Conclusion}

%Our work aims at contributing to the development of formal models
%suitable for describing biological phenomena and automatically
%reasoning about their properties. In this paper we have focused on
%causality in metabolic networks. We have suggested how detailing the
%structure of a causal relation: {\em how} a metabolite can be
%produced adds expressiveness to reasoning about {\em what} can be
%produced. In this context we have discussed some properties, largely
%used by biologists, made formal. We have introduced a computational
%tool supporting the verification of the properties, which is based on
%techniques developed for dealing with causality in concurrency. Finally,
%we have outlined some simple case studies suggesting the relevance of
%the approach and links with possible application domains.

%Future directions for this approach regard a systematic formalization of
%as many as possible properties of biological interest, and the
%development of computational tools for addressing large scale {\em in
%silico} experiments for metabolic networks.

\scriptsize
\bibliographystyle{eptcs}
\bibliography{BBCG_FBTC10.bib}

\begin{thebibliography}{10}
\providecommand{\bibitemstart}[1]{\bibitem{#1}}
\providecommand{\bibitemend}{}
\providecommand{\bibliographystart}{}
\providecommand{\bibliographyend}{}
\providecommand{\url}[1]{\texttt{#1}}
\providecommand{\urlprefix}{Available at }
\providecommand{\bibinfo}[2]{#2}
\bibliographystart

\bibitemstart{AJB00}
\bibinfo{author}{R.~Albert}, \bibinfo{author}{H.~Jeong} \&
  \bibinfo{author}{A.~Barabasi} (\bibinfo{year}{2000}):
  \emph{\bibinfo{title}{Error and attack tolerance of complex networks}}.
\newblock {\sl \bibinfo{journal}{Nature}} \bibinfo{volume}{406}.
\bibitemend

\bibitemstart{Alon07}
\bibinfo{author}{U.~Alon} (\bibinfo{year}{2007}): \emph{\bibinfo{title}{Network
  motifs: theory and experimental approaches}}.
\newblock {\sl \bibinfo{journal}{Nature reviews Genetics}} \bibinfo{volume}{8}.
\bibitemend

\bibitemstart{shank}
\bibinfo{author}{C.~Baral}, \bibinfo{author}{W.~Kolch},
  \bibinfo{author}{C.~Shankland} \& \bibinfo{author}{N.~Tran}
  (\bibinfo{year}{2005}): \emph{\bibinfo{title}{Reasoning about the ERK signal
  transduction pathway using BioSigNet-RR}}.
\newblock In: {\sl \bibinfo{booktitle}{Proc.~of CMSB'05}}.
\bibitemend

\bibitemstart{B09}
\bibinfo{author}{C.~Bodei} (\bibinfo{year}{2009}): \emph{\bibinfo{title}{A
  Control Flow Analysis for Beta-binders with and without static
  compartments}}.
\newblock {\sl \bibinfo{journal}{Theoretical Computer Science}}
  \bibinfo{volume}{410}(\bibinfo{number}{33-34}).
\bibitemend

\bibitemstart{BMC08}
\bibinfo{author}{C.~Bodei}, \bibinfo{author}{A.~Bracciali} \&
  \bibinfo{author}{D.~Chiarugi} (\bibinfo{year}{2008}):
  \emph{\bibinfo{title}{On deducing causality in metabolic networks}}.
\newblock {\sl \bibinfo{journal}{BMC Bioinformatics}} \bibinfo{volume}{9 (4)}.
\bibitemend

\bibitemstart{brane}
\bibinfo{author}{L.~Cardelli} (\bibinfo{year}{2005}):
  \emph{\bibinfo{title}{Brane calculi}}.
\newblock In: {\sl \bibinfo{booktitle}{Proc.~of Computational Methods in
  Systems Biology (CMSB'04)}},  \bibinfo{volume}{LNCS 3082}.
  \bibinfo{organization}{Springer}, pp. \bibinfo{pages}{257--280}.
\bibitemend

\bibitemstart{Ca08}
\bibinfo{author}{L.~Cardelli} (\bibinfo{year}{2008}): \emph{\bibinfo{title}{On
  process rate semantics}}.
\newblock {\sl \bibinfo{journal}{Theoretical Computer Science}}
  \bibinfo{volume}{391}(\bibinfo{number}{3}), pp. \bibinfo{pages}{190--215}.
\bibitemend

\bibitemstart{CFS05}
\bibinfo{author}{N.~Chabrier}, \bibinfo{author}{F.~Fages} \&
  \bibinfo{author}{S.~Soliman} (\bibinfo{year}{2005}):
  \emph{\bibinfo{title}{The Biochemical abstract machine BIOCHAM}}.
\newblock In: {\sl \bibinfo{booktitle}{Proc.~of CMSB'04}},
  \bibinfo{volume}{LNCS 3082}. \bibinfo{organization}{Springer}, pp.
  \bibinfo{pages}{172--191}.
\bibitemend

\bibitemstart{Clyde}
\bibinfo{author}{R.G. Clyde}, \bibinfo{author}{J.L. Bown},
  \bibinfo{author}{T.R. Hupp}, \bibinfo{author}{N.~Zhelev} \&
  \bibinfo{author}{J.W. Crawford} (\bibinfo{year}{2006}):
  \emph{\bibinfo{title}{The role of modelling in identifying drug targets for
  diseases of the cell cycle}}.
\newblock {\sl \bibinfo{journal}{J R Soc Interface}} \bibinfo{volume}{3}, pp.
  \bibinfo{pages}{617--627}.
\bibitemend

\bibitemstart{CGL08}
\bibinfo{author}{A.~Coletta}, \bibinfo{author}{R.~Gori} \&
  \bibinfo{author}{F.~Levi} (\bibinfo{year}{2009}):
  \emph{\bibinfo{title}{Approximating Probabilistic Behaviors of Biological
  Systems Using Abstract Interpretation}}.
\newblock In: {\sl \bibinfo{booktitle}{Proc.~of FBTC'08}},
  \bibinfo{volume}{ENTCS 229(1)}. pp. \bibinfo{pages}{165--182}.
\bibitemend

\bibitemstart{CDPB04}
\bibinfo{author}{M.~Curti}, \bibinfo{author}{P.~Degano},
  \bibinfo{author}{C.~Priami} \& \bibinfo{author}{C.T. Baldari}
  (\bibinfo{year}{2004}): \emph{\bibinfo{title}{Modeling biochemical pathways
  through enhanced pi-calculus}}.
\newblock {\sl \bibinfo{journal}{Theoretical Computer Science}}
  \bibinfo{volume}{325}(\bibinfo{number}{1}), pp. \bibinfo{pages}{111--140}.
\bibitemend

\bibitemstart{rccs}
\bibinfo{author}{V.~Danos} \& \bibinfo{author}{J.~Krivine}
  (\bibinfo{year}{2005}): \emph{\bibinfo{title}{Transactions in RCCS}}.
\newblock In: {\sl \bibinfo{booktitle}{Proc.~of CONCUR'05}},
  \bibinfo{volume}{LNCS 3653}. \bibinfo{organization}{Springer}.
\bibitemend

\bibitemstart{kappa}
\bibinfo{author}{V.~Danos} \& \bibinfo{author}{C.~Laneve}
  (\bibinfo{year}{2003}): \emph{\bibinfo{title}{Graphs for Core Molecular
  Biology}}.
\newblock In: {\sl \bibinfo{booktitle}{Proc.~of CMSB'03}},
  \bibinfo{volume}{LNCS 2602}. \bibinfo{organization}{Springer}, pp.
  \bibinfo{pages}{34 -- 46}.
\bibitemend

\bibitemstart{EKLLT02}
\bibinfo{author}{S.~Eker}, \bibinfo{author}{M.~Knapp},
  \bibinfo{author}{P.~Lincoln}, \bibinfo{author}{K.~Laderoute} \&
  \bibinfo{author}{C.~Talcott} (\bibinfo{year}{2002}):
  \emph{\bibinfo{title}{Pathway Logic: Executable Models of Biological
  Network}}.
\newblock In: {\sl \bibinfo{booktitle}{Proc.~of the Fourth International
  Workshop on Rewriting Logic and Its Applications (WRLA'02), ENTCS 71}}.
  \bibinfo{organization}{Elsevier}, pp. \bibinfo{pages}{144--161}.
\bibitemend

\bibitemstart{Fatumo}
\bibinfo{author}{S.~Fatumo}, \bibinfo{author}{K.~Plaimas},
  \bibinfo{author}{J.P. Mallm}, \bibinfo{author}{G.~Schramm},
  \bibinfo{author}{E.~Adebiyi}, \bibinfo{author}{M.~Oswald},
  \bibinfo{author}{R.~Eils} \& \bibinfo{author}{R.~König}
  (\bibinfo{year}{2009}): \emph{\bibinfo{title}{Estimating novel potential drug
  targets of Plasmodium falciparum by analysing the metabolic network of
  knock-out strains in silico}}.
\newblock {\sl \bibinfo{journal}{Infect. Genet. Evol.}} \bibinfo{volume}{9(3)},
  pp. \bibinfo{pages}{351--8}.
\bibitemend

\bibitemstart{Fell97}
\bibinfo{author}{D.A. Fell} (\bibinfo{year}{1997}):
  \emph{\bibinfo{title}{Understanding the control of metabolism}}.
\newblock \bibinfo{publisher}{Portland Press, London, United Kingdom}.
\bibitemend

\bibitemstart{Gambhir}
\bibinfo{author}{S.S Gambhir} (\bibinfo{year}{2002}):
  \emph{\bibinfo{title}{Molecular imaging of cancer with positron emission
  tomorgraphy}}.
\newblock {\sl \bibinfo{journal}{Nat. Rev. Cancer}} \bibinfo{volume}{2}, pp.
  \bibinfo{pages}{891--899}.
\bibitemend

\bibitemstart{Gatenby}
\bibinfo{author}{R.A. Gatenby} \& \bibinfo{author}{R.J. Gillies}
  (\bibinfo{year}{2007}): \emph{\bibinfo{title}{Glycolysis in cancer: A
  potential target for therapy}}.
\newblock {\sl \bibinfo{journal}{The International Journal of Biochemistry and
  Cell Biology}} \bibinfo{volume}{39}, pp. \bibinfo{pages}{1358--1366}.
\bibitemend

\bibitemstart{Gerdes}
\bibinfo{author}{S.~Gerdes}, \bibinfo{author}{R.~Edwards},
  \bibinfo{author}{M.~Kubal}, \bibinfo{author}{M.~Fonstein},
  \bibinfo{author}{R.~Stevens} \& \bibinfo{author}{A.~Osterman}
  (\bibinfo{year}{2006}): \emph{\bibinfo{title}{Essential genes on metabolic
  maps}}.
\newblock {\sl \bibinfo{journal}{Current Opinion in Biotechnology}}
  \bibinfo{volume}{17}, pp. \bibinfo{pages}{448--456}.
\bibitemend

\bibitemstart{Gill00}
\bibinfo{author}{G.T. Gillespie} (\bibinfo{year}{2000}):
  \emph{\bibinfo{title}{The chemical Langevin equation}}.
\newblock {\sl \bibinfo{journal}{Journal of Chemical Physics}}
  \bibinfo{volume}{113}(\bibinfo{number}{1}), pp. \bibinfo{pages}{297--306}.
\bibitemend

\bibitemstart{GLi&c09}
\bibinfo{author}{R.~Gori} \& \bibinfo{author}{F.~Levi} (\bibinfo{year}{2009}):
  \emph{\bibinfo{title}{Abstract Interpretation based Verification of Temporal
  Properties for BioAmbients.}} \bibinfo{note}{To appear in Info \&Co.}
\bibitemend

\bibitemstart{GL09}
\bibinfo{author}{R.~Gori} \& \bibinfo{author}{F.~Levi} (\bibinfo{year}{2009}):
  \emph{\bibinfo{title}{Abstract Interpretation for Probabilistic Termination
  of Biological Systems.}}
\newblock In: {\sl \bibinfo{booktitle}{Proc.~of MeCBIC'09}},
  \bibinfo{volume}{EPTCS 11}.
\bibitemend

\bibitemstart{K02}
\bibinfo{author}{H~Kitano} (\bibinfo{year}{2002}):
  \emph{\bibinfo{title}{Systems Biology: a brief overview}}.
\newblock {\sl \bibinfo{journal}{Science}}
  \bibinfo{volume}{295}(\bibinfo{number}{5560}), pp.
  \bibinfo{pages}{1662--1664}.
\bibitemend

\bibitemstart{Kitanob}
\bibinfo{author}{H.~Kitano} (\bibinfo{year}{2004}):
  \emph{\bibinfo{title}{Biological robustness}}.
\newblock {\sl \bibinfo{journal}{Nat Rev Genet}} \bibinfo{volume}{5}, pp.
  \bibinfo{pages}{826--837}.
\bibitemend

\bibitemstart{Kitanoa}
\bibinfo{author}{H.~Kitano} (\bibinfo{year}{2007}):
  \emph{\bibinfo{title}{Towards a theory of biological robustness}}.
\newblock {\sl \bibinfo{journal}{Molecular Systems Biology}}
  \bibinfo{volume}{3}, p. \bibinfo{pages}{137}.
\bibitemend

\bibitemstart{Klamt}
\bibinfo{author}{S.~Klamt} \& \bibinfo{author}{E.D. Gilles}
  (\bibinfo{year}{2002}): \emph{\bibinfo{title}{Minimal cut sets in biochemical
  reaction networks}}.
\newblock {\sl \bibinfo{journal}{Nature}} \bibinfo{volume}{420}.
\bibitemend

\bibitemstart{pi}
\bibinfo{author}{R.~Milner} (\bibinfo{year}{1999}):
  \emph{\bibinfo{title}{Communicating and mobile systems: the $\pi$-calculus}}.
\newblock \bibinfo{publisher}{Cambridge University Press}.
\bibitemend

\bibitemstart{CFA_BIO}
\bibinfo{author}{F.~Nielson}, \bibinfo{author}{H.~Riis-Nielson},
  \bibinfo{author}{D.~Schuch-Da-Rosa} \& \bibinfo{author}{C.~Priami}
  (\bibinfo{year}{2004}): \emph{\bibinfo{title}{Static analysis for systems
  biology}}.
\newblock In: {\sl \bibinfo{booktitle}{Proc.~of workshop on
  Systeomatics-dynamic biological systems informatics}}.
  \bibinfo{publisher}{Computer Science Press, Trinity College Dublin,}, pp.
  \bibinfo{pages}{1--6}.
\bibitemend

\bibitemstart{cfaBioAmb}
\bibinfo{author}{F.~Nielson}, \bibinfo{author}{H.~Riis-Nielson},
  \bibinfo{author}{D.~Schuch-Da-Rosa} \& \bibinfo{author}{C.~Priami}
  (\bibinfo{year}{2007}): \emph{\bibinfo{title}{Control Flow Analysis for
  BioAmbients}}.
\newblock {\sl \bibinfo{journal}{ENTCS}}
  \bibinfo{volume}{180}(\bibinfo{number}{3}), pp. \bibinfo{pages}{65--79}.
\bibitemend

\bibitemstart{Palsson}
\bibinfo{author}{Y-K. Oh}, \bibinfo{author}{B.Ø. Palsson},
  \bibinfo{author}{S.M. Park}, \bibinfo{author}{C.H. Schilling} \&
  \bibinfo{author}{R.~Mahadevan} (\bibinfo{year}{2007}):
  \emph{\bibinfo{title}{Genome-scale Reconstruction of Metabolic Network in
  Bacillus subtilis Based on High-throughput Phenotyping and Gene Essentiality
  Data}}.
\newblock {\sl \bibinfo{journal}{TJ Biol Chem}} \bibinfo{volume}{10}, pp.
  \bibinfo{pages}{693--706}.
\bibitemend

\bibitemstart{Orosz}
\bibinfo{author}{F.~Orosz}, \bibinfo{author}{J.~Oláh} \&
  \bibinfo{author}{J.~Ovádi} (\bibinfo{year}{2006}):
  \emph{\bibinfo{title}{Triosephosphate isomerase deficiency: facts and
  doubts.}}
\newblock {\sl \bibinfo{journal}{IUBMB life}} \bibinfo{volume}{12}, pp.
  \bibinfo{pages}{703--715}.
\bibitemend

\bibitemstart{PC04}
\bibinfo{author}{A.~Phillips} \& \bibinfo{author}{L.~Cardelli}
  (\bibinfo{year}{2004}): \emph{\bibinfo{title}{A correct abstract machine for
  the stochastic pi-calculus}}.
\newblock In: {\sl \bibinfo{booktitle}{Proc.~of Bioconcur'04}}.
  \bibinfo{organization}{ENTCS, Elsevier.}
\bibitemend

\bibitemstart{newcfaBioAmb}
\bibinfo{author}{H.~Pilegaard}, \bibinfo{author}{F.~Nielson} \&
  \bibinfo{author}{H.~Riis Nielson} (\bibinfo{year}{2008}):
  \emph{\bibinfo{title}{Pathway analysis for BioAmbients}}.
\newblock {\sl \bibinfo{journal}{J. Log. Algebr. Program.}}
  \bibinfo{volume}{77}(\bibinfo{number}{1-2}), pp. \bibinfo{pages}{92--130}.
\bibitemend

\bibitemstart{cfaBioAmbLDL}
\bibinfo{author}{H.~Pilegaard}, \bibinfo{author}{H.~Riis Nielson} \&
  \bibinfo{author}{F.~Nielson} (\bibinfo{year}{2006}):
  \emph{\bibinfo{title}{Static Analysis of a Model of the LDL Degradation
  Pathway}}.
\newblock In: {\sl \bibinfo{booktitle}{Simulation and Verification of Dynamic
  Systems}}.
\bibitemend

\bibitemstart{Pr05}
\bibinfo{author}{C.~Priami} (\bibinfo{year}{1995}):
  \emph{\bibinfo{title}{Stochastic $\pi$-calculus}}.
\newblock {\sl \bibinfo{journal}{The Computer Journal}} \bibinfo{volume}{38},
  pp. \bibinfo{pages}{578--589}.
\bibitemend

\bibitemstart{beta1}
\bibinfo{author}{C.~Priami} \& \bibinfo{author}{P.~Quaglia}
  (\bibinfo{year}{2005}): \emph{\bibinfo{title}{Beta Binders for Biological
  Interactions}}.
\newblock In: {\sl \bibinfo{booktitle}{Proc.~of CMSB'04}},
  \bibinfo{volume}{LNCS 3082}. \bibinfo{organization}{Springer}, pp.
  \bibinfo{pages}{20--33}.
\bibitemend

\bibitemstart{PRSS04}
\bibinfo{author}{C.~Priami}, \bibinfo{author}{A.~Regev},
  \bibinfo{author}{E.~Shapiro} \& \bibinfo{author}{W.~Silvermann}
  (\bibinfo{year}{2001}): \emph{\bibinfo{title}{Application of a stochastic
  name-passing calculus to representation and simulation of molecular
  processes}}.
\newblock {\sl \bibinfo{journal}{Inf. Process. Lett.}}
  \bibinfo{volume}{80}(\bibinfo{number}{1}), pp. \bibinfo{pages}{25--31}.
\bibitemend

\bibitemstart{RPS+04}
\bibinfo{author}{A.~Regev}, \bibinfo{author}{E.~Panina},
  \bibinfo{author}{W.~Silverman}, \bibinfo{author}{L.~Cardelli} \&
  \bibinfo{author}{E.~Shapiro} (\bibinfo{year}{2004}):
  \emph{\bibinfo{title}{Bioambients: An abstraction for biological
  compartements}}.
\newblock {\sl \bibinfo{journal}{Theoretical Computer Science}}
  \bibinfo{volume}{325}(\bibinfo{number}{1}), pp. \bibinfo{pages}{141--167}.
\bibitemend

\bibitemstart{bio_pi}
\bibinfo{author}{A.~Regev}, \bibinfo{author}{W.~Silvermann} \&
  \bibinfo{author}{E.~Shapiro} (\bibinfo{year}{2001}):
  \emph{\bibinfo{title}{Representation and Simulation of Biochemical Processes
  Using the pi-Calculus Process Algebra}}.
\newblock In: {\sl \bibinfo{booktitle}{Pacific Symposium on Biocomputing}}. pp.
  \bibinfo{pages}{459--470}.
\bibitemend

\bibitemstart{Schilling}
\bibinfo{author}{C.H. Schilling}, \bibinfo{author}{D.~Letscher} \&
  \bibinfo{author}{B.O. Palsson} (\bibinfo{year}{2000}):
  \emph{\bibinfo{title}{Theory for the systemic definition of metabolic
  pathways and their use in interpreting metabolic function from a
  pathway-oriented perspective}}.
\newblock {\sl \bibinfo{journal}{J. Theor. Biol.}} \bibinfo{volume}{203}.
\bibitemend

\bibitemstart{Schuster}
\bibinfo{author}{S.~Schuster}, \bibinfo{author}{D.~Fell} \&
  \bibinfo{author}{T.~Dandekar} (\bibinfo{year}{2000}): \emph{\bibinfo{title}{A
  general definition of metabolic pathways useful for systematic organization
  and analysis of complex metabolic networks}}.
\newblock {\sl \bibinfo{journal}{Nat. Biotechnol}} \bibinfo{volume}{18}, pp.
  \bibinfo{pages}{226--232}.
\bibitemend

\bibitemstart{Stelling}
\bibinfo{author}{S.~Schuster}, \bibinfo{author}{D.~Fell} \&
  \bibinfo{author}{T.~Dandekar} (\bibinfo{year}{2002}):
  \emph{\bibinfo{title}{Metabolic network structure determines key aspects of
  functionality and regulation}}.
\newblock {\sl \bibinfo{journal}{Nature}} \bibinfo{volume}{420}, pp.
  \bibinfo{pages}{190--193}.
\bibitemend

\bibitemstart{SSEB03}
\bibinfo{author}{B.~Shargel}, \bibinfo{author}{H.~Sayama},
  \bibinfo{author}{I.R. Epstein} \& \bibinfo{author}{Y.~Bar-Yam}
  (\bibinfo{year}{2003}): \emph{\bibinfo{title}{Optimization of Robustness and
  Connectivity in Complex Networks}}.
\newblock {\sl \bibinfo{journal}{Physical Review Letters}}
  \bibinfo{volume}{90}.
\bibitemend

\bibitemstart{Stellingj}
\bibinfo{author}{J.~Stelling}, \bibinfo{author}{U.~Sauer},
  \bibinfo{author}{Z.~Szallasi}, \bibinfo{author}{F.J.~Doyle III} \&
  \bibinfo{author}{J.~Doyle} (\bibinfo{year}{2004}):
  \emph{\bibinfo{title}{Robustness of cellular functions}}.
\newblock {\sl \bibinfo{journal}{Cell}} \bibinfo{volume}{118}, pp.
  \bibinfo{pages}{675--685}.
\bibitemend

\bibitemstart{TN03}
\bibinfo{author}{J.~Thykaer} \& \bibinfo{author}{J.~Nielsen}
  (\bibinfo{year}{2003}): \emph{\bibinfo{title}{Metabolic engineering of
  beta-lactam production.}}
\newblock {\sl \bibinfo{journal}{Metabolic Eng.}} \bibinfo{volume}{5 (1)}.
\bibitemend

\bibitemstart{Yu}
\bibinfo{author}{B.J. Yu}, \bibinfo{author}{B.H. Sung}, \bibinfo{author}{J.Y.
  Lee}, \bibinfo{author}{S.H. Son}, \bibinfo{author}{M.S. Kim} \&
  \bibinfo{author}{S.C. Kim} (\bibinfo{year}{2006}):
  \emph{\bibinfo{title}{sucAB and sucCD are mutually essential genes in
  Escherichia coli}}.
\newblock {\sl \bibinfo{journal}{FEMS Microbiol Lett}} \bibinfo{volume}{254},
  pp. \bibinfo{pages}{245--250}.
\bibitemend

\bibitemstart{YK07}
\bibinfo{author}{H.~Yu}, \bibinfo{author}{P.M. Kim},
  \bibinfo{author}{E.~Sprecher}, \bibinfo{author}{V.~Trifonov} \&
  \bibinfo{author}{M.~Gerstein} (\bibinfo{year}{2009}):
  \emph{\bibinfo{title}{The importance of bottlenecks in protein networks:
  correlation with gene essentiality and expression dynamics.}}
\newblock {\sl \bibinfo{journal}{PLOS Computational Biology}}
  \bibinfo{volume}{3 (4)}.
\bibitemend

\bibliographyend
\end{thebibliography}

\end{document}